\documentclass[a4paper,fleqn]{cas-dc}
\usepackage[T1]{fontenc} 
\usepackage[utf8]{inputenc}
\usepackage[english]{babel}
\usepackage[numbers,sort&compress]{natbib}
\usepackage[most]{tcolorbox}
\usepackage{subcaption}
\usepackage{threeparttable}
\usepackage{latexsym} 
\usepackage{arydshln} 

\graphicspath{{figs/}}

\newcounter{suggestions}
\newtcolorbox{suggestionbox}{
    enhanced,
    breakable,
    sharp corners,
    boxrule=0.3pt,
    colback=white,
    colframe=black!40,
    coltitle=black,
    fonttitle=\bfseries,
    toptitle=2mm,
    bottomtitle=2mm,
    colbacktitle=white,
    title={Suggestion~\#\refstepcounter{suggestions}\thesuggestions}}

\newtcolorbox{summarybox}{
    enhanced,
    unbreakable,
    sharp corners,
    colback=white,
    colframe=black!60,
    coltitle=black}
    
\usepackage[indentfirst=false,font={itshape},begintext=``,endtext='']{quoting}
\usepackage{epigraph}

\def\indie{Indie\xspace}
\def\nonindie{Non-indie\xspace}

\def\tsc#1{\csdef{#1}{\textsc{\lowercase{#1}}\xspace}}
\tsc{WGM}
\tsc{QE}
\tsc{EP}
\tsc{PMS}
\tsc{BEC}
\tsc{DE}

\begin{document}

\let\WriteBookmarks\relax
\def\floatpagepagefraction{1}
\def\textpagefraction{.001}

\title[mode = title]{Game Industry Problems: an Extensive Analysis of the Gray Literature}
\shorttitle{Game Industry Problems: an Extensive Analysis of the Gray Literature}

\author[1]{Cristiano Politowski}[orcid=0000-0002-0206-1056] \cormark[1]
\ead{c_polito@encs.concordia.ca}

\author[2]{Fabio Petrillo}[orcid=0000-0002-8355-1494]
\ead{fabio@petrillo.com}

\author[3]{Gabriel C. Ullmann}
\ead{gabriel.cavalheiro@sou.unijui.edu.br}

\author[1]{Yann-Ga\"el Gu\'{e}h\'{e}neuc}[orcid=0000-0002-4361-2563]
\ead{yann-gael.gueheneuc@concordia.ca}

\address[1]{Concordia University, Montreal, Quebec, Canada}
\address[2]{Université du Québec à Chicoutimi, Chicoutimi, Quebec, Canada}
\address[3]{Universidade Regional do Noroeste do Estado do Rio Grande do Sul, Santa Rosa, Rio Grande do Sul, Brazil}

\cortext[cor1]{Corresponding author}

\shortauthors{Politowski, Petrillo, Ullmann, and Gu\'{e}h\'{e}neuc.}

\newcommand{\projectpage}{\url{https://github.com/game-dev-database/postmortem-problems}}
\newcommand{\totalGroups}{4}
\newcommand{\totalTypes}{20}
\newcommand{\totalSubtypes}{105}
\newcommand{\totalProblems}{927}

\newcommand{\todo}[1]{\textcolor{blue}{TODO: #1} }
\newcommand{\fabio}[1]{\textcolor{red}{$>>>$ Fabio: #1 $<<<$}}
\newcommand{\YANN}[1]{\textcolor{red}{$>>>$ Yann: #1 $<<<$}}
\newcommand{\cris}[1]{\textcolor{red}{$>>>$ Cris: #1 $<<<$}}

\def\indie{Indie\xspace}
\def\nonindie{Non-indie\xspace}

\begin{abstract}
Context: Given its competitiveness, the video-game industry has a closed-source culture. Hence, little is known about the problems faced by game developers. However, game developers do share information about their game projects through postmortems, which describe informally what happened during the projects.
Objective: The software-engineering research community and game developers would benefit from a state of the problems of the video game industry, in particular the problems faced by game developers, their evolution in time, and their root causes. This state of the practice would allow researchers and practitioners to work towards solving these problems.
Method: We analyzed 200 postmortems from 1997 to 2019, resulting in 927 problems divided into 20 types. Through our analysis, we described the overall landscape of game industry problems in the past 23 years and how these problems evolved over the years. We also give details on the most common problems, their root causes, and possible solutions. We finally discuss suggestions for future projects.
Results: We observe that 
(1) the game industry suffers from management and production problems in the same proportion;
(2) management problems decreased over the years, giving space to business problems, while production problems remained constant;
(3a) technical and game design problems are decreasing over the years, the latter only after the last decade;
(3b) problems related to the team increase over the last decade;
(3c) marketing problems are the ones that had the biggest increase over the 23 years compared to other problem types;
(4) finally, the majority of the main root causes are related to people, not technologies.
Conclusions: In this paper, we provide a state of the practice for researchers to understand and study video-game development problems. We also offer suggestions to help practitioners to avoid the most common problems in future projects.
\end{abstract}


\maketitle






\section{Introduction}

\epigraph{``\textit{The history of science, like the history of all human ideas, is a history of irresponsible dreams, of obstinacy, and of error}''}{Karl Popper}


\paragraph{Context:} As technology evolves, it offers improved video-game experiences that attract more and more players\footnote{\url{https://review42.com/video-game-statistics/}}, therefore making video-games the most profitable entertainment industry sector today\footnote{\url{https://bit.ly/30fSjLq}}.

The game industry is known for its problems. They range from technical problems, e.g., 80\% of the games on Steam require critical updates \cite{Lin2016}, to management problems, e.g., crunch time \cite{Edholm2017} and unrealistic scopes \cite{Petrillo2009}. The problems in the game industry also include mistreatment of employees\footnote{\url{https://bit.ly/3h6ZKer}} and harassment\footnote{\url{https://bit.ly/391zf7B}}. Yet, the game industry continues to make profits\footnote{\url{https://bit.ly/3haHukG}} as players keep on buying its games, reinforcing a cycle of bad practices.

These high-profile problems are possibly only the proverbial tip of the iceberg. Indeed, little is known of the problems faced day-to-day by game developers. Game studios are secretive and have a closed-source culture. Although not exclusive to the game industry, closed-source projects are preponderant in the game industry, while open-source projects are more common in traditional software development. Open-source games are rare, and the main tools used by game developers, i.e., game engines, are proprietary (e.g. Unity and Unreal). For example, GitHub has few popular game-engine projects while its counterparts, software frameworks, are numerous \cite{Politowski2020}.

Yet, contrary to other software industries, game developers \emph{do} share information about their games projects in the form of ``war stories''. These war stories are postmortems, which are informal texts that summarise the developers' experiences with their games projects, often written by managers or senior developers \cite{Callele2005} right after their games launched \cite{Washburn2016}. They often include sections about ``What went right'' and ``What went wrong'' during the game development:

\begin{itemize}
\item \emph{``What went right''} discusses the best practices adopted by the game developers, solutions, improvements, and project-management decisions that helped the project.
	
\item \emph{``What went wrong''} discusses difficulties, pitfalls, and mistakes experienced by the development team in the project, both technical and managerial.
\end{itemize}

\paragraph{Objective:} Game developers and the software-engineering research community would benefit from a state of the problems of video-game development, in particular the problems faced by game developers, their evolution in time, and their root causes. This state of the practice would allow researchers and practitioners to work towards solving these problems.

\paragraph{Method:} We analyse 200 postmortems written between 1997 and 2019 available in our public dataset \cite{Politowski2020dataset} of grey literature related to game development. These postmortems include \totalProblems~problems that we categorized into \totalTypes~types. Through our analysis, we draw a landscape of game-industry problems in the past 23 years and how these problems evolved over the years. We give details on the most common problems, their root causes, and possible solutions. We also provide suggestions for future projects.

\paragraph{Results:} For each of the \totalProblems~problems in the dataset, we identify its \emph{root causes} and \emph{solutions}. We show that:

\begin{itemize}
\item Based on the number of problems groups and types, the game industry suffers from \textit{management} and \textit{production} problems in the same proportion. However,  \textit{production} problems are concentrated mostly in \textit{technical} and \textit{design} problems while \textit{management} problems are more distributed across problem types.

\item Based on the evolution of problem groups over the years, \textit{management} problems decreased, giving space to \textit{business} problems, while \textit{production} problems remained constant;

\item Based on the evolution of the problem types over the years:
    \begin{itemize}
    \item \textit{Technical} and \textit{game design} problems are decreasing, the latter only in the last decade;
    
    \item Problems related to \textit{teams} increased in the last decade;
    
    \item \textit{Marketing} problems have the greatest increase, over the 23 years, compared to other problem types;
    \end{itemize}

\item Considering the problem sub-types, the main root causes are related to people, not technologies.
\end{itemize}

\paragraph{Conclusions:} This analysis describes the problems faced by game developers during their game projects and some of their solutions. It shows that many problems require project-specific solutions that are hard to generalise, while others do not yet have clearly defined answers. Thus, we provide a state of the practice for researchers to understand and study video-game development problems. We also offer suggestions to help practitioners to avoid the most common problems in future projects.

The paper is structured as follows:  Section~\ref{sec:dataset} describes the dataset. Section~\ref{sec:res-birdeye} shows the overall analysis of the problems and its evolution over the years.
Section~\ref{sec:res-subtypes} further discusses the top 10 root causes and outlines the developers' solutions.
Section~\ref{sec:related} presents the related work and compares their findings with our data.
Section~\ref{sec:discussion} discusses our suggestions for the top 10 problems.
Section~\ref{sec:threats} describes the threats to validity. Section~\ref{sec:conclusion} concludes the paper with future work.

\section{Dataset} \label{sec:dataset}

For this study, we extended the dataset defined in the previous work \cite{Politowski2020dataset}.
In this Section, we summarize how we created, analysed, and expanded this dataset.

\subsection{Method}

\begin{figure}[hbt!]
	\centering
	\includegraphics[width=1\linewidth]{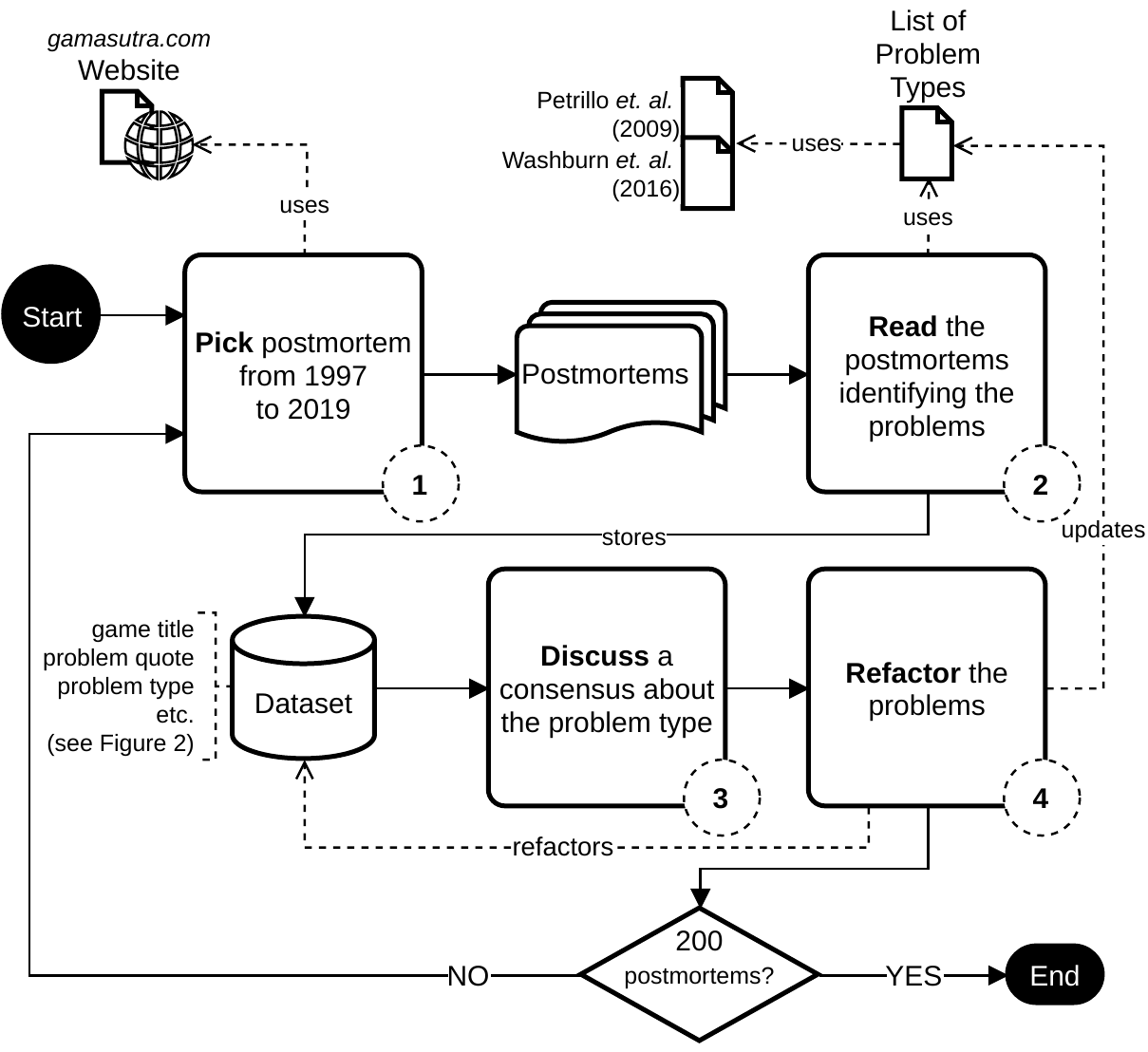}
	\caption{Steps performed to analyse the postmortems and build the dataset.}
	\label{fig:method}
\end{figure}

Our analysis process was iterative, where the data from the postmortems was constantly evolving, allowing extracted refactoring in each new iteration. \autoref{fig:method} shows the process of collecting and compiling the data from the postmortems.

\paragraph{Step 1:}
We started with each author randomly picking one postmortem from the Gamasutra Website between the years 1997 to 2019. 

\paragraph{Step 2:}
Each author read the postmortem, focusing on the ``What went wrong'' section. Using the coding technique from Grounded Theory \cite{Stol2016}, we identified the problems reported by the game developers, extracting \textit{quotes}, and grouping similar \textit{problem types}. 
As a starting point, we created a list of \textit{problem types} based on the previous literature definitions of \citet{Petrillo2009} and \citet{Washburn2016}. 

\paragraph{Step 3:}
In this step we discussed the findings, and reached a consensus about the \textit{problem types}.

\paragraph{Step 4:}
At this step, any change resulted in updates in the \textit{dataset} and the list of \textit{problem types}. This process continued until we reached 200 postmortems.

At approximately 140 postmortem's reviews, the list of \textit{problem types} stabilized, i.e., the problems we found could be allocated to the existing types.
Also, to keep the distribution of the postmortems temporally balanced, some of them were manually chosen instead of randomly picked. Other postmortems were replaced when they did not contain useful information regarding game development. Therefore, reviewed reading more than 200 postmortems.

\subsection{Dataset Metadata}
\label{sec:metadata}


The list of \emph{problem types} is a document listing the problems found in the postmortems. First, it included only the problem types gathered from the literature \cite{Petrillo2009, Washburn2016}. We updated it with every newly problem type discovered, after the authors reached a consensus about the new problem type. \autoref{tab:list-problem-types} shows the final version of the catalogue.

\begin{table}[pos=htb!]
\footnotesize
\caption{%
\textit{List of problem types} of video-game development problems identified through the postmortem analysis. 
The \textit{types} that are also used by \citet{Petrillo2009} or \citet{Washburn2016} are described \textit{P}, and \textit{W}, respectively.}
\label{tab:list-problem-types}
\begin{tabular*}{\tblwidth}{@{} p{\dimexpr.27\linewidth}  p{\dimexpr.68\linewidth} @{}}
\toprule
Type & Description \\ \midrule
Bugs\textsuperscript{P} & Bugs or failures that compromise the game development or its reception. \\ \addlinespace
Game Design\textsuperscript{PW} & Game design problems, like balancing the gameplay, creating fun mechanics, etc. \\ \addlinespace
Documentation\textsuperscript{PW} & Not documenting the code, artifacts or game plan. \\ \addlinespace
Prototyping & Lack of or no prototyping phase nor validation of the gameplay/feature. \\ \addlinespace
Technical\textsuperscript{P} & Problems with code or assets, infra-structure, network, hardware, etc. \\ \addlinespace
Testing\textsuperscript{PW} & Any problem regarding testing the game, like unit tests, playtesting, QA, etc. \\ \addlinespace
Tools\textsuperscript{PW} & Problems with tools like Game Engines, libraries, etc. \\ \midrule
Communication\textsuperscript{P} & Problems communicating with any stakeholder, team, publisher, audience, etc. \\ \addlinespace
Crunch Time\textsuperscript{P} & When developers continuously spent extra hours working in the project. \\ \addlinespace
Delays & Problems regarding any delay in the project. \\ \addlinespace
Team\textsuperscript{PW} & Problems in setting up the team, loss of professionals during the development or outsourcing. \\ \addlinespace \midrule
Cutting Features\textsuperscript{P} & Cutting features previously planned due to other factors like time or budget. \\ \addlinespace
Feature Creep\textsuperscript{P} & Adding non-planned new features to the game during its production. \\ \addlinespace
Multiple Projects & When there is more than one project being developed at the same time. \\ \addlinespace
Budget\textsuperscript{PW} & Lack of budget, funding, and any financial difficulties. \\ \addlinespace
Planning\textsuperscript{W} & Problems involving planning and schedule, or lack of either. \\ \addlinespace
Security & Problems regarding leaked assets or information about the project. \\ \addlinespace
Scope\textsuperscript{PW} & When the project is has too many features that end up impossible to implement it. \\ \midrule
Marketing\textsuperscript{W} & Problems regarding marketing and advertising. \\ \addlinespace
Monetization & Problems with the process used to generate revenue from a video game product. \\ \bottomrule
\end{tabular*}
\end{table}



To have a better macro idea of the problems, we decreased the granularity of the problem types by clustering them into four groups:
\begin{itemize}
    \item \textit{Production} describes practical problems that often happen during the production phase; 
    \item \textit{People Management} describes management problems related to people; 
    \item \textit{Feature Management} describes management problems related to the game features; and, 
    \item \textit{Business} describes the marketing and the strategy to generate revenue.
\end{itemize}


On the other hand, to further investigate the root causes of the problems, we continued increasing the granularity by re-reading the problems and classifying them with a more specific description called \textit{problem sub-types}.
For example, \autoref{tab:dataset-structure} shows one  dataset entry (one problem). The game ``Baldur's Gate II'' from 2001 was analyzed and it has a problem \textit{type} ``Testing'', which belongs to the \textit{group} ``Production''. After the second analysis we defined the \textit{sub-type} as ``Scope too big to test properly''. The \textit{quote} is an excerpt from the postmortem.

\begin{table}[pos=htb!]
	\footnotesize
	\caption{Example of one entry in the dataset.}
	\label{tab:dataset-structure}
	\begin{tabular*}{\tblwidth}{@{}>{\raggedleft\arraybackslash}p{\dimexpr.15\linewidth} p{\dimexpr.8\linewidth} @{}}
		\toprule
		Column & Value \\
		\midrule
		ID & 61 \\
		Title & Baldurs Gate II -- The Anatomy of a Sequel \\
		Year & 2001 \\
		Source & \url{http://bit.ly/2IDsVa0} \\
		Name & Baldur's Gate II \\
		Platform & PC \\
		Genre & RPG Strategy \\
		Mode & Multi Single \\
		\textbf{Group} & Production \\
		\textbf{Type} & Testing \\
		\textbf{SubType} & Scope too big to test properly \\
		\textbf{Quote} & 
		(...) We put a number of white-boards in the halls of the testing and design area and listed all of the quests on the boards. We then put an X next to each quest. We broke the designers and QA teams into paired subgroups - each pair (one tester and one designer) had the responsibility of thoroughly checking and fixing each quest. After they were certain the quest was bulletproof, its X was removed. It took about 2 weeks to clear the board (on the first pass). \\
		\bottomrule
	\end{tabular*}
\end{table}

\begin{figure}[htb!]
\centering
\includegraphics[width=1\linewidth]{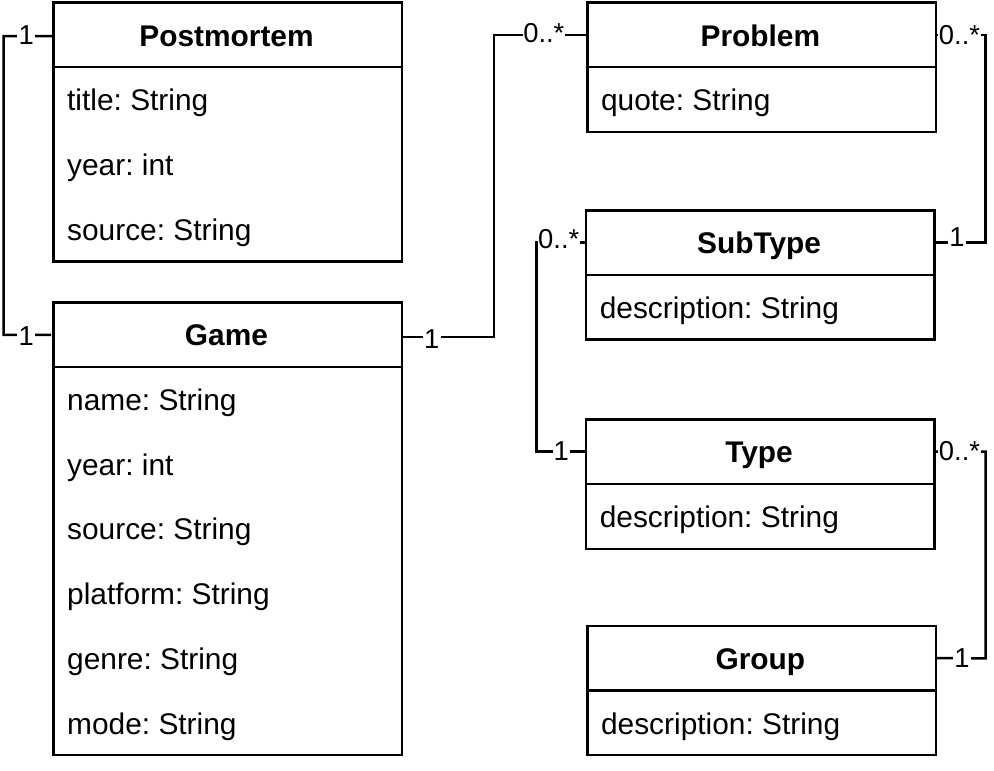}
\caption{Class diagram with the structure of each entry on the dataset.}
\label{fig:structure}
\end{figure}

To store the problems gathered from postmortems, we defined a data model. \autoref{fig:structure} shows its UML class diagram. Each \textit{Postmortem} relates to one \textit{Game} as the document describes what happened in only one project. The \textit{Game} has a collection of \textit{Problems}. With increasing granularity, each \textit{Problem} has a \textit{SubType}, a \textit{Type}, and a \textit{Group}. 

A \textit{Game} also has: \textit{Platform} [1-3] (PC, Console, Mobile), \textit{Genre} [1-12] (Action, Adventure, RPG, Simulation, Strategy, Puzzle, Sports, Platformer, Shooter, Racing, Roguelike, Running\footnote{This is a short list of the most common game genres: \url{https://en.wikipedia.org/wiki/Video_game_genre}.}), \textit{Mode} [1-3] (Single/Multi Player, Online).

The dataset is available in a open repository on GitHub\footnote{\projectpage} so that researchers and practitioners can access and contribute through  \emph{pull requests}. We choose this approach to curate contributions before inclusion. Contributors can also add problem types and other metadata, e.g., genres, to the list.
\section{Overview of the Problem Types}
\label{sec:res-birdeye}

This Section shows the results of the dataset analysis. Section~\ref{sec:res-overall} describes the dataset, the problem groups and types and problems by platform (PC, Console, Mobile). Section~\ref{sec:res-years} shows the evolution of problems over the years.

\subsection{Overall Dataset}
\label{sec:res-overall}

\begin{figure}[pos=htb!]
\centering
\begin{subfigure}{.8\linewidth}
\includegraphics[width=1\linewidth]{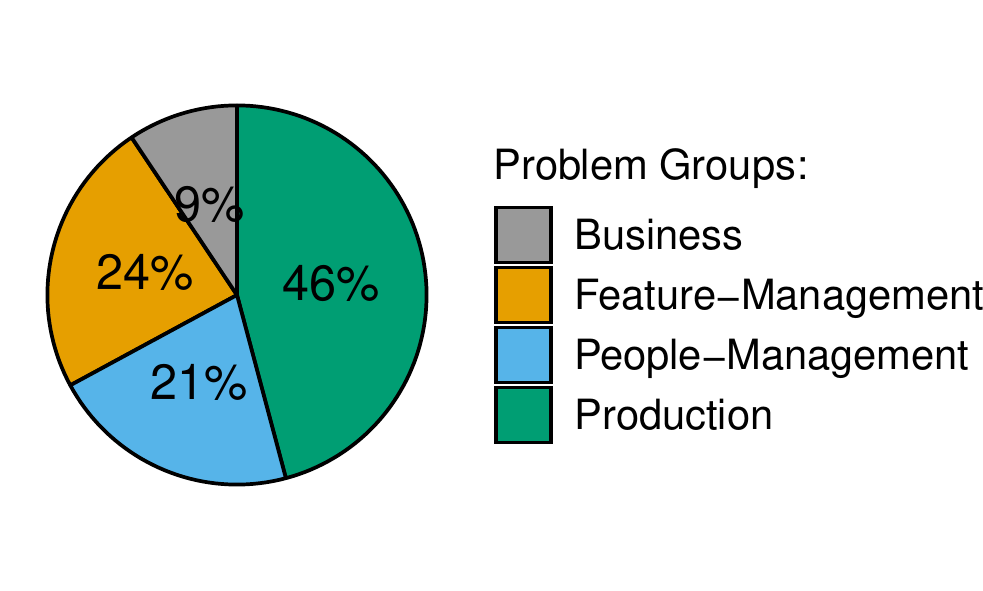}
\caption{Number of problems related to each Group.}
\label{fig:problem-group}
\end{subfigure}

\begin{subfigure}{1\linewidth}
\includegraphics[width=1\linewidth]{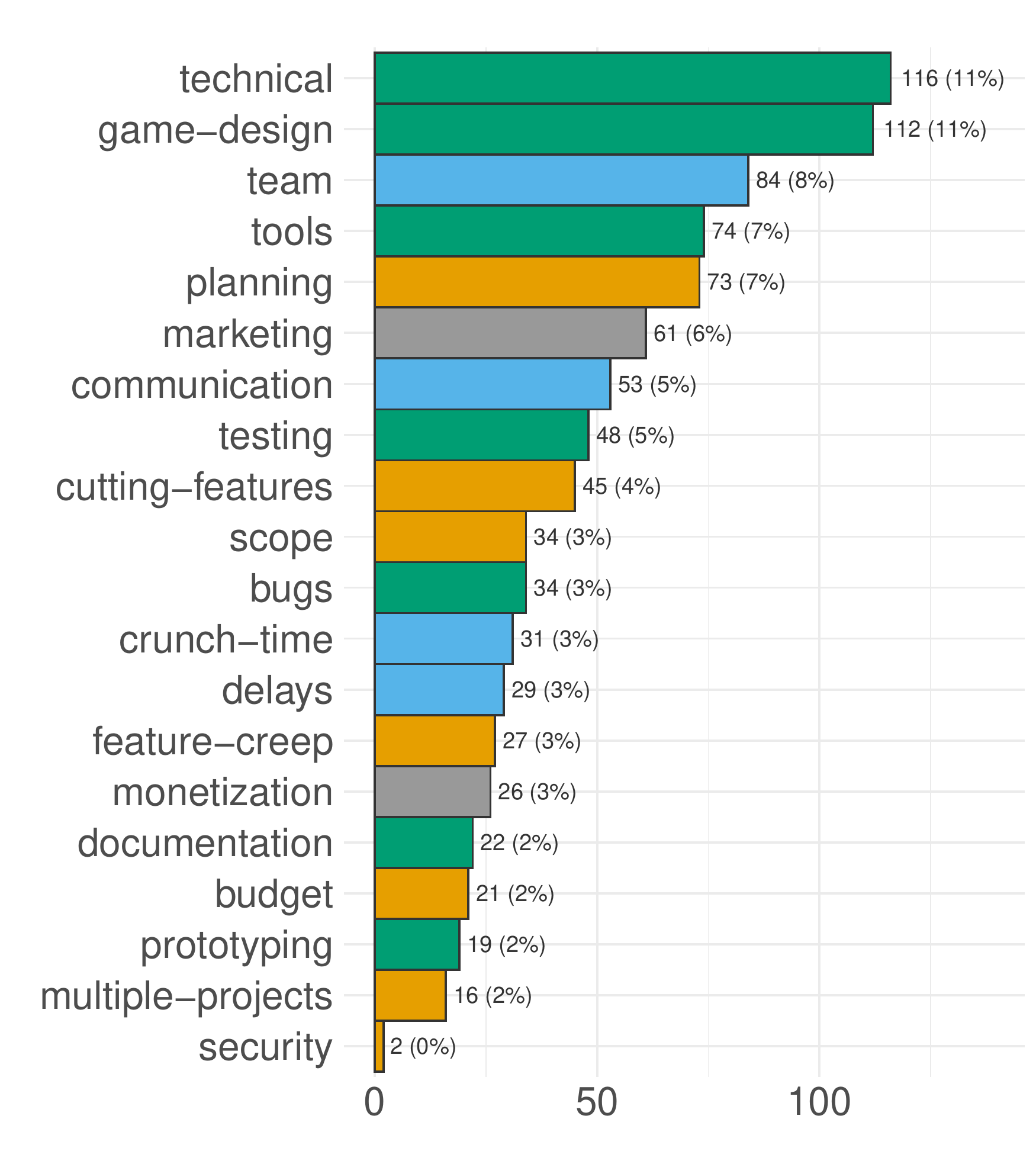}
\caption{Number of problems related to each Type.}
\label{fig:problem-type}
\end{subfigure}
\caption{Overall dataset results for problem groups and types.}
\label{fig:problem-type-group}
\end{figure}

The dataset contains 200 video-game projects from 1997 to 2019, describing \totalProblems~problems. On average, there are five problems by game and 40 by year. \autoref{fig:problem-group} shows the problems by groups: 46\% of the problems relate to production, 45\% to management, and 9\% to business.

\autoref{fig:problem-type} shows the distribution of the problems by types. \textit{Game design}, \textit{technical}, and \textit{team} problems are the most frequent, with 30\% overall. Although management and production problems have close percentages, the two most common problems types, \textit{technical} and \textit{game design}, with 11\% each, are related to \textit{production}. \textit{Management} problems are spread among problem types.

\begin{figure}[htb!]
	\centering
	\includegraphics[width=.8\linewidth]{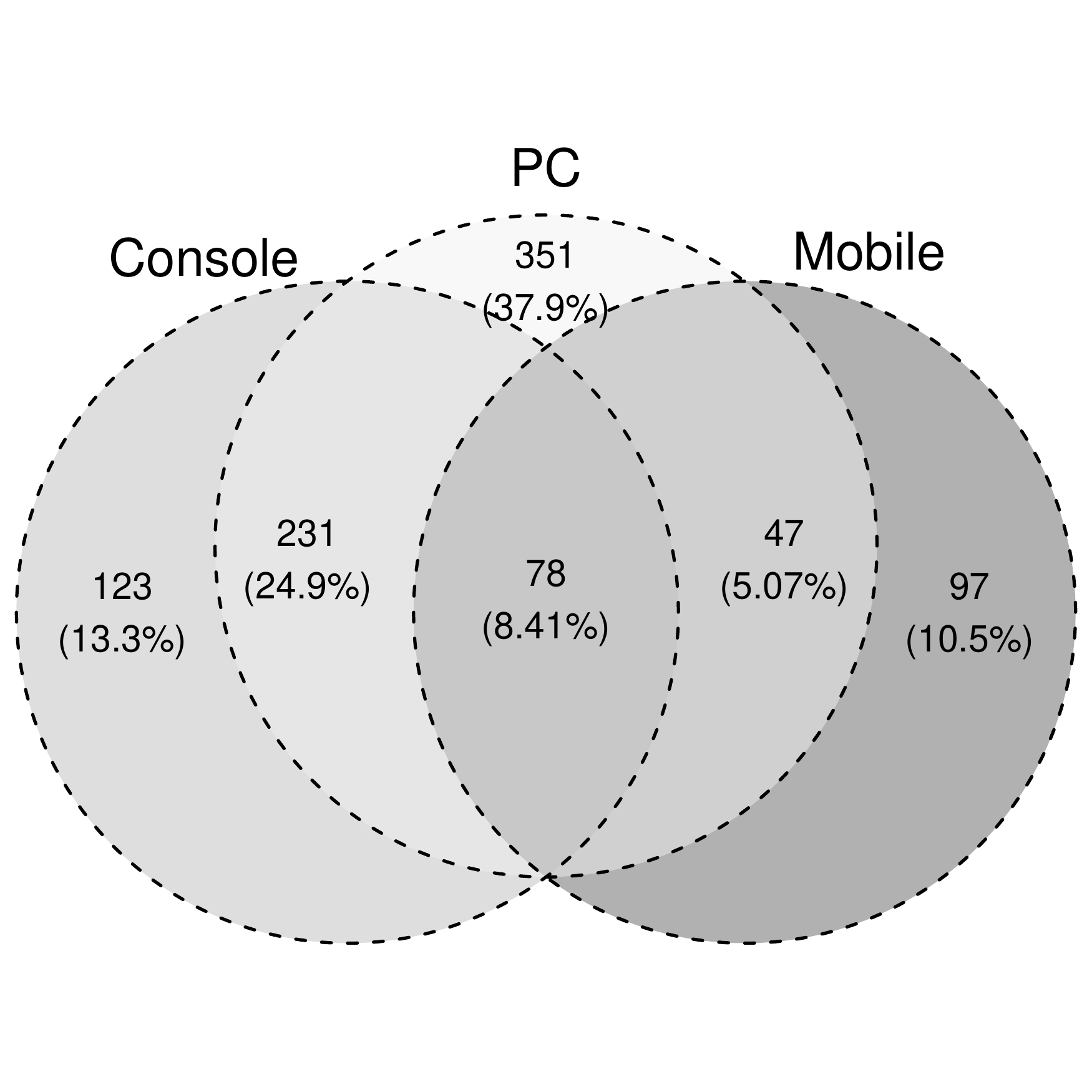}
	\caption{Venn-diagram of the problems by platforms: PC, Console, and Mobile.}
	\label{fig:problems-platforms}
\end{figure}

\autoref{fig:problems-platforms} shows the problems by game platforms: PC, Console, and Mobile. The problems described in the postmortems mainly occur in PC games, with 707 problems, followed by 432 Console problems, and 222 Mobile problems. Only 78 problems pertain to the three platforms (multi-platform games). Mobile and PC games are more likely to be ported to Consoles. No game was made only for Mobile and Console without also being on PC.

\subsection{Problems Over the Years}
\label{sec:res-years}

\begin{figure}[htb!]
	\centering
	\includegraphics[width=1\linewidth]{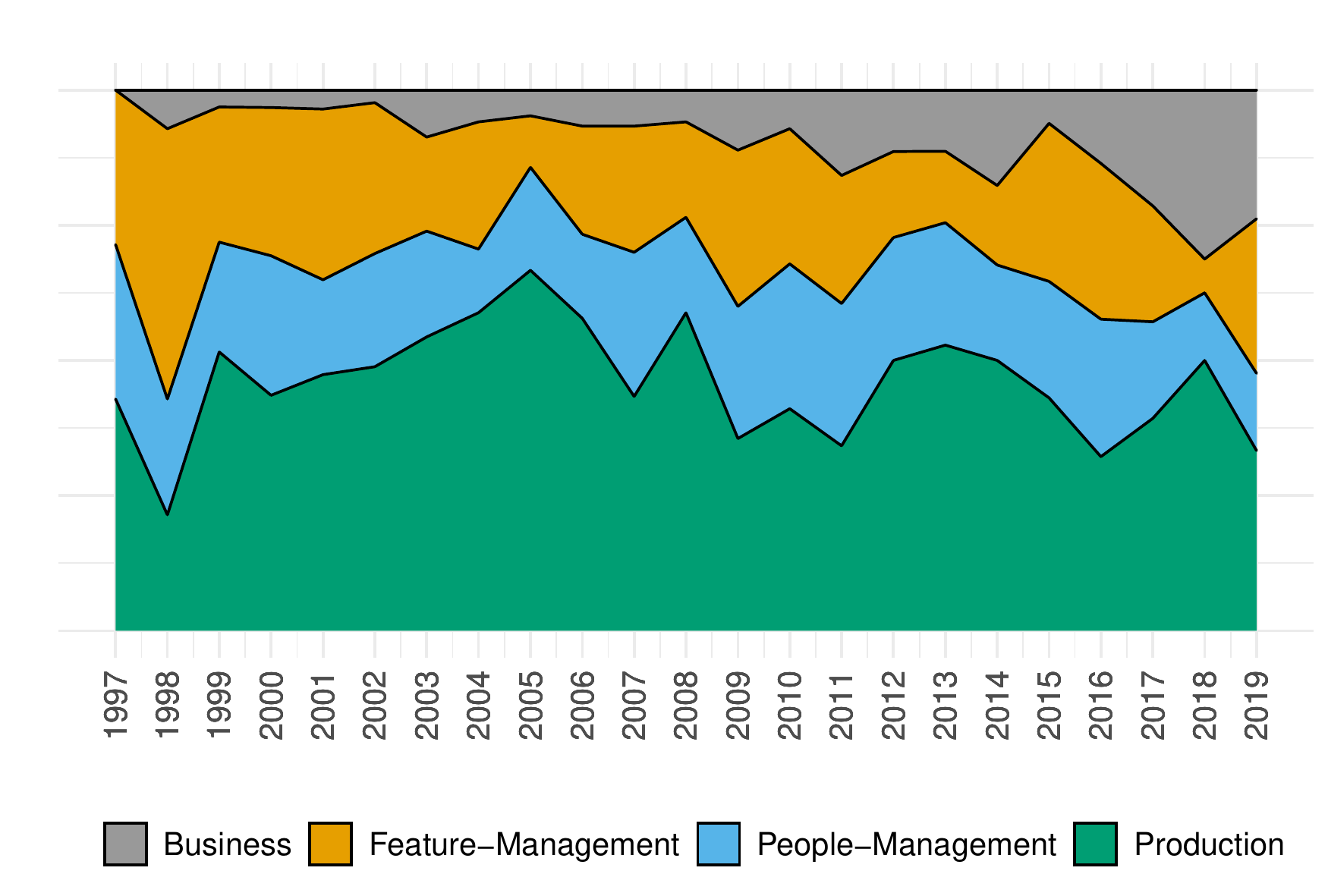}
	\caption{Problems over the years by groups.}
	\label{fig:hist-group}
\end{figure}

\autoref{fig:hist-group} shows the normalised number of problems per group or per year. For example, in 2018, there were five business problems among 16 problems. \textit{Production} problems remain constant. \textit{Management} problems peaked in 1998 and are less frequent now. \textit{Business} problems increased over the years.

\begin{figure*}[width=1\textwidth,pos=htb!]
\centering
\begin{subfigure}{.48\linewidth}
	\includegraphics[width=1\linewidth]{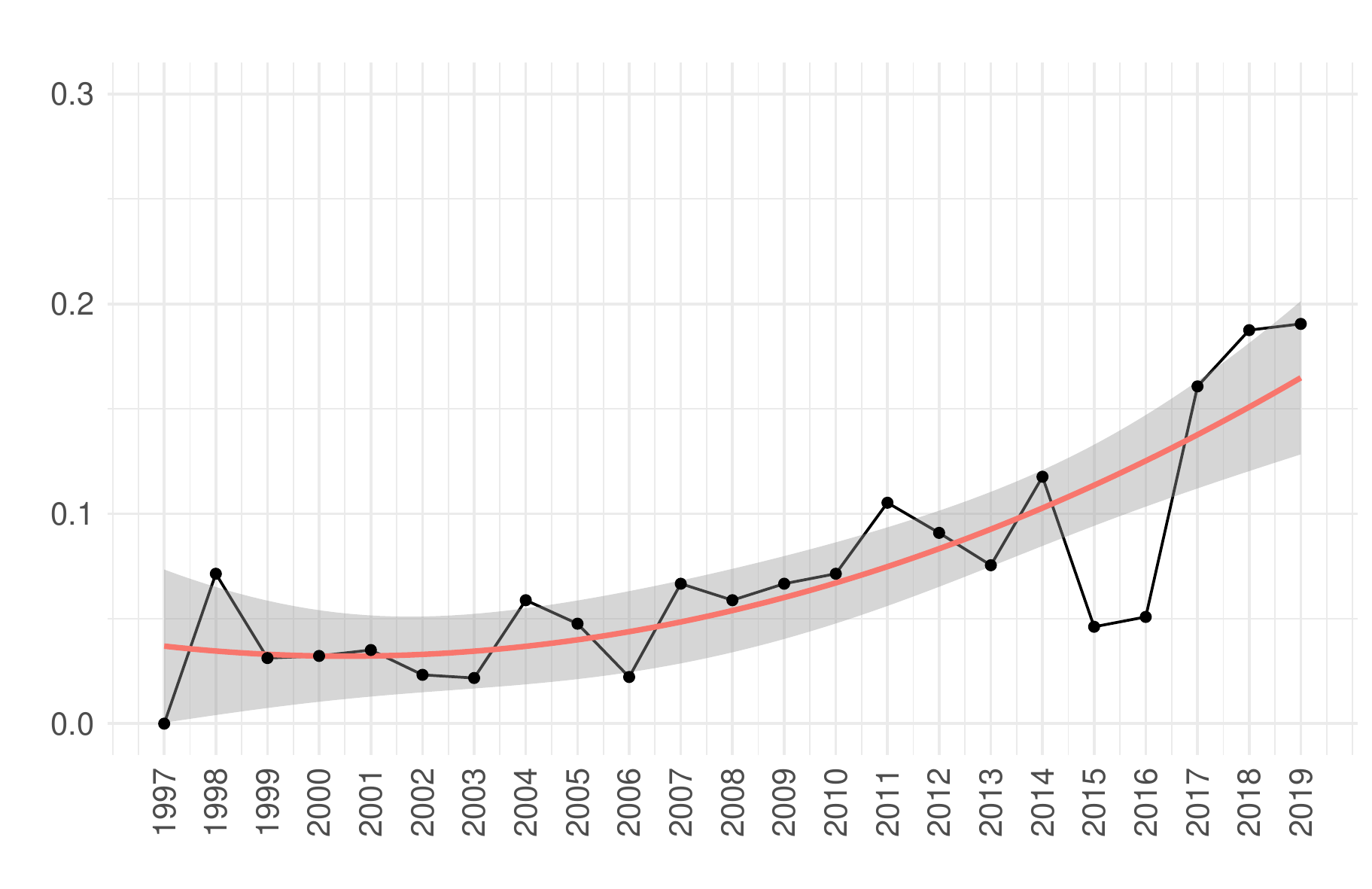}
	\caption{Marketing problems}
	\label{fig:historical-patterns-marketing}
\end{subfigure}
\begin{subfigure}{.48\linewidth}
	\includegraphics[width=1\linewidth]{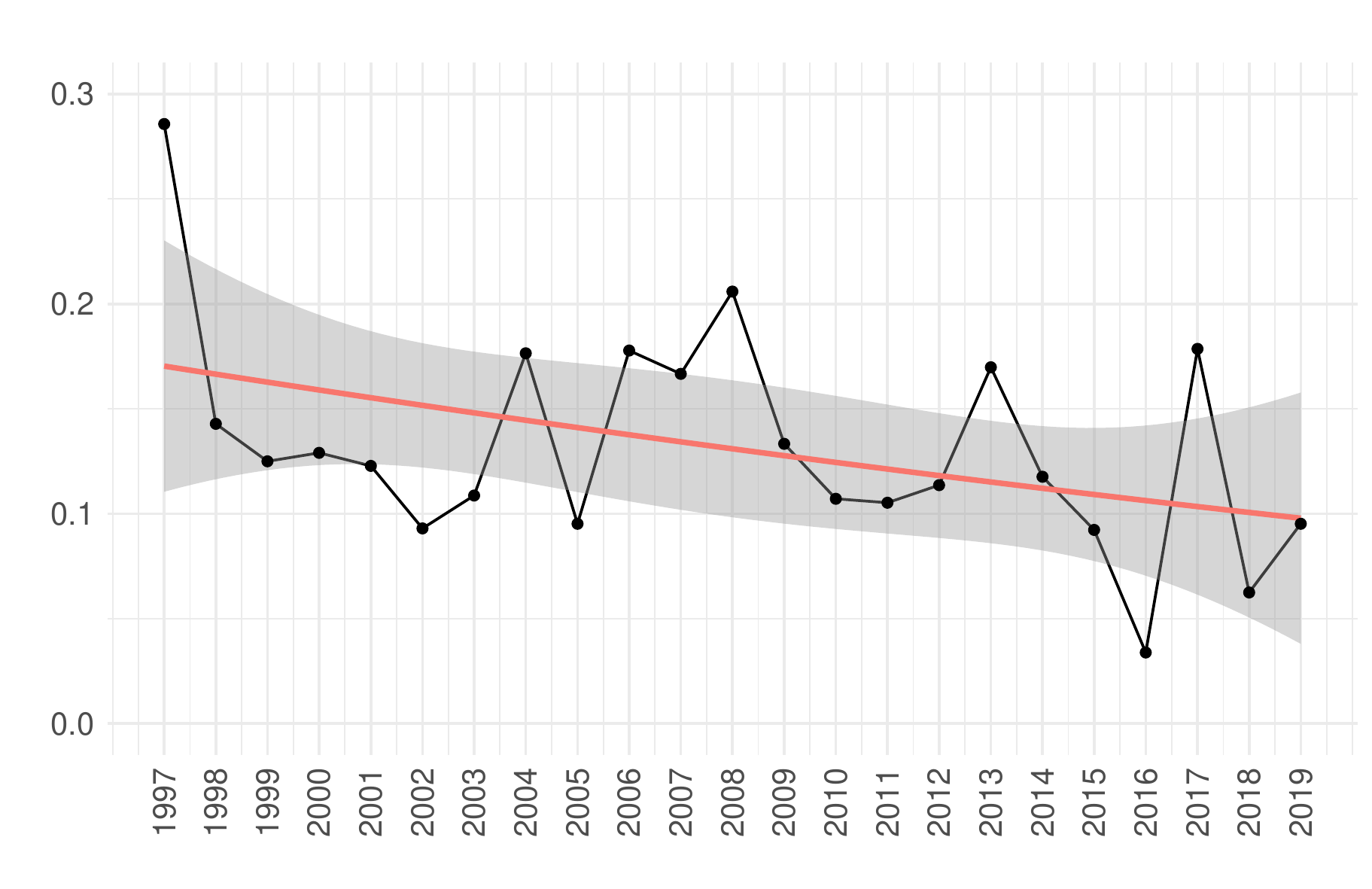}
	\caption{Technical problems}	
	\label{fig:historical-patterns-technical}
\end{subfigure}

\begin{subfigure}{.48\linewidth}
	\includegraphics[width=1\linewidth]{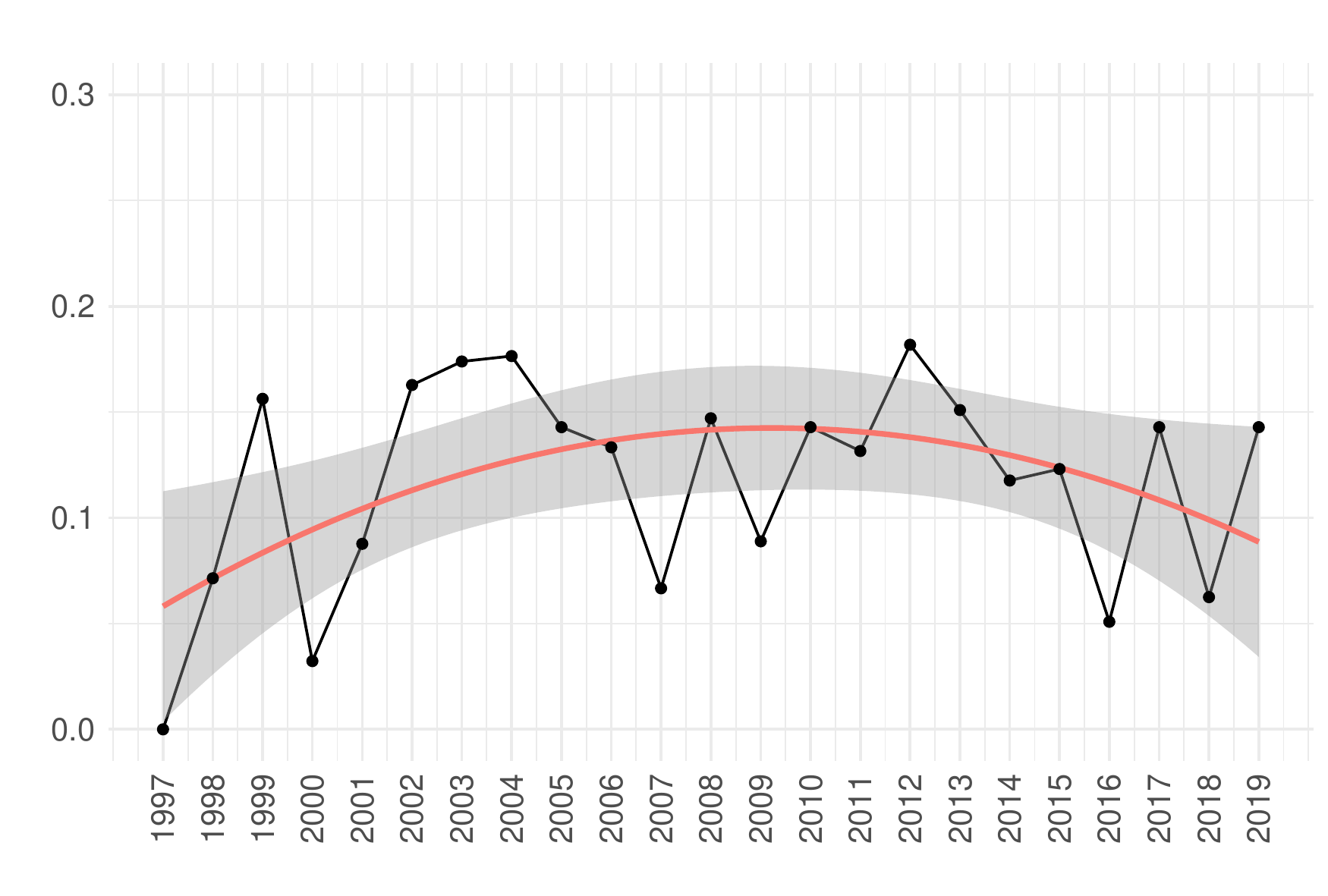}
	\caption{Game Design problems}
	\label{fig:historical-patterns-design}
\end{subfigure}
\begin{subfigure}{.48\linewidth}
	\includegraphics[width=1\linewidth]{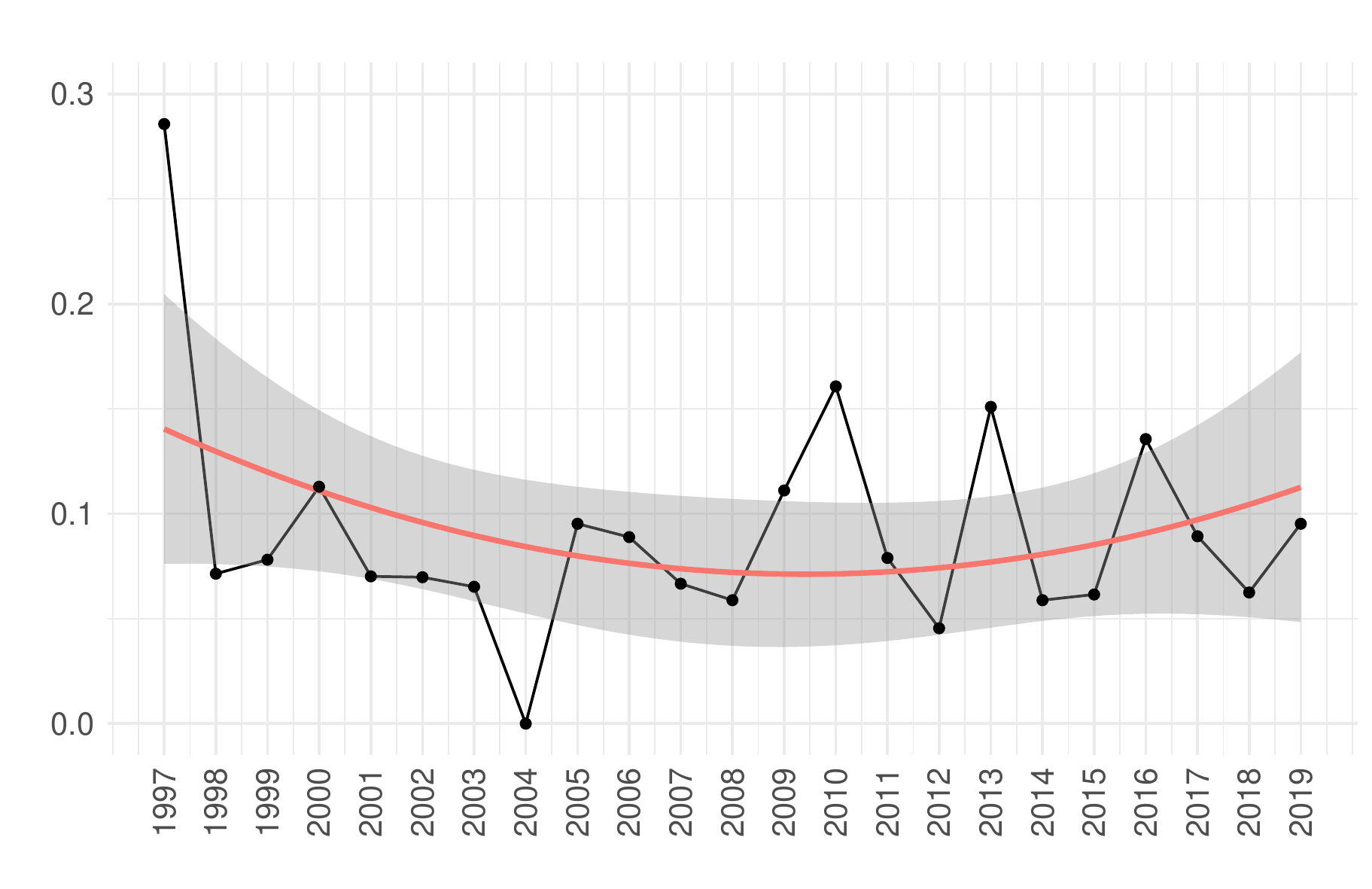}
	\caption{Team problems}
	\label{fig:historical-patterns-team}
\end{subfigure}

\caption{Four common patterns of the importance of the problems over the years. (a) Shows Marketing problem that increased since 1997. (b) Shows Technical problems that decreased since 1997. (c) Shows Game Design problems that decreased in the last decade. (d) Shows Team problems that increased in the last decade.}
\label{fig:historical-patterns}
\end{figure*}

\autoref{fig:historical-patterns} shows the four different patterns in the dataset. To normalise the numbers of problems each year, we divide their numbers by the total numbers of problems that year. The red line (curved line) is a second-degree polynomial function. The grey area represents the confidence interval (0.95 by default) of the function.

\autoref{fig:historical-patterns-marketing} shows that \textit{Marketing} problems increase over the years. \textit{Monetization} and \textit{Bugs} are also problems that follow this trend, but to a lesser degree.

On the contrary, \autoref{fig:historical-patterns-technical} shows the decrease of \textit{Technical} problems over the years. Other problem types also follow this trend: \textit{Documentation}, \textit{Testing}, \textit{Cutting Features}, and to a lesser degree \textit{Feature Creep} and \textit{Communication}.

We also observe problem types whose trends changed in the last decade. For example, \autoref{fig:historical-patterns-design} shows that \textit{Game Design} problems, the most notorious case of a problems, decreased in the last decade. To a lesser degree, this pattern is followed by the problem types: \textit{Tools}, \textit{Delays}, \textit{Crunch Time}, \textit{Budget}, \textit{Planning}, and \textit{Prototyping}.

However, some problems increased in the last decade. \autoref{fig:historical-patterns-team} shows the most evident example of problems related to development \textit{Teams}. Problems related to project \textit{Scope} also follow this trend.

\section{Top Ten Problem Sub-types}
\label{sec:res-subtypes}

\newcommand{\explanation}{Explanation:}
\newcommand{\solutions}{Solutions:}

In this section, we further investigate the problems and identify the root causes of each problem type. We read all the problems again classifying the types into sub-types. We found a total of \totalSubtypes~different sub-types. \autoref{tab:subtypes} describes the top 10 sub-type problems. We focus on the top 10 sub-types for lack of space while \autoref{sec:app-subtypes} presents all the sub-types. In the following, each subsection discusses one problem sub-type. It provides an \textit{Explanation} and proposed \textit{Solutions} to the problem sub-type, illustrated by excerpts from postmortems. All the information in this Section is based on the postmortems.

\begin{table}[pos=htb!]
\caption{The top 10 most common sub-type problems.}
\label{tab:subtypes}
\begin{tabular}{@{}lll@{}}
\toprule
Type          & SubType (root cause)                                                 & N  \\ \midrule
Team          & \hyperref[sec-understaffing]{Insufficient workforce}             & 49 \\
Team          & \hyperref[sec-environment]{Environment problems} & 48 \\
Marketing     & \hyperref[sec-marketing]{Wrong marketing strategy}                                & 35 \\
Planning      & \hyperref[sec-understimation]{Underestimation}                                         & 34 \\
Game Design        & \hyperref[sec-vision]{Unclear game design vision}                                   & 28 \\
Game Design        & \hyperref[sec-fun]{Lack of fun}                                   & 27 \\
Technical     & \hyperref[sec-platform]{Platform and technology constraints}                     & 24 \\
Game Design        & \hyperref[sec-complexity]{Game design complexity}                    & 23 \\
Tools         & \hyperref[sec-tools]{Inadequate or missing tools}                              & 22 \\
Communication & \hyperref[sec-team]{Misaligned teams}                                    & 22 \\
\bottomrule
\end{tabular}
\end{table}

\subsection{Insufficient Workforce}
\label{sec-understaffing}

\paragraph{\explanation} Insufficient workforce is the main issue among the team problems. It happens when a game company does not have enough developers for all the tasks or when a developer has too many tasks. This problem often happens when there is inadequate budget for the game project. Other causes include a lack of planning (tasks, schedules, testing, etc), the difficulty to find developers with certain skill sets, experience, and willingness to work on a game project.

\begin{quoting}[endtext='' -- P\#604]
For the first six months of production, one person was juggling design, project management, and a number of significant project-external responsibilities. They were -- obviously -- overtasked. It led to a lack of communication on scheduling between studio management and the development team. 
\end{quoting}

\paragraph{\solutions} ``Hire more, share the load'' is the most frequent advice given by developers. They also mention outsourcing and remote work as solutions to the budget constraints. Other alternatives to mitigate this problem include dividing the tasks among more people to improve efficiency, calling for help and staffing up sooner when needed. Moreover, be sure to dismiss the outsourced professionals only when their job is 100\% complete and integrated in the game.

\subsection{Environmental Problems}
\label{sec-environment}

\paragraph{\explanation} Even a properly staffed and well experienced team may suffer if their corporate environment has problems. Environments, especially in large studios, are a source of problems when, for example, there is a lack of a departmental organisation or hierarchy. Low wages, lack of incentives, toxic behaviours (e.g., harassment or bullying), excessive or mandatory crunch times, lack of open communication, and lack of working standards are all environmental problems.

\begin{quoting}[endtext='' -- P\#467]
We would've been better off had we realized that personality fit and talent aren't enough: people need to mesh with your working style. When people are unhappy, it spreads through the whole team.
\end{quoting}

\paragraph{\solutions} Avoiding environmental problems in game studios requires balancing the developers' experience and having experienced developers mentoring junior developers. It also includes keeping the team cohesive, decentralising decisions, putting the ``right'' developer on the ``right'' task, and having small sub-teams with smaller scopes. It also involves supporting the team with material and psychological resources for morale. Finally, it helps to have a fixed, but spaced, meeting dates, especially when following a well-defined process with a clear hierarchy.

\subsection{Wrong Marketing Strategy}
\label{sec-marketing}


\paragraph{\explanation} Although game developers are not necessarily competent at marketing, they may overlook its importance as they are focused on finishing the game. Developers commonly, and wrongly, try reaching a broad audience by giving away copies of their games to specialized media and demoing their games in game conferences. According to the developers, this strategy is not effective.

Related problems include targeting the ``wrong'' audience (for example, sending a copy of a RPG game to a person that reviews racing games), miscommunication with the players, promoting the game too much, and losing marketing opportunities (e.g: major sales holidays such as Christmas). Developers also reported specific marketing problems in crowdfunding campaigns and in early-access programs.

\begin{quoting}[endtext='' -- P\#430]
The launch week I started looking at [YouTube] streamers for the PC version, so I basically searched for big youtubers that covered games like: Spelunky, Meat boy, and a few other more recent pixel-art indie games that fit the same category as [the game]. I mailed all of them, close to a 100, with at least one steam-key included (...) and this all resulted in an awesome 0 [streams]. I did a follow up email to a large portion of them a week later, and this resulted in 1 Streamer playing it, yay results! 
\end{quoting}

\paragraph{\solutions} ``Don't announce until you're much closer to release'' is the most common advice from the developers.
Game studios often announce their games years in advance hoping to ``build the hype'' and get noticed. Developers also recommend working with the players, invest time producing marketing material, and, if possible, producing a demo of the game to create awareness and build a community. They also warn not to ``oversell'' (promise too much) games and focus on the game development, in particular if the game studio does not have marketing expertise. They also suggest launching games in more than one store/marketplace, producing the launch trailers, and focusing the marketing on the games strengths. Finally, they recommend distributing a limited and traced number of game copies to reviewers to avoid piracy.

\subsection{Underestimation}
\label{sec-understimation}

\paragraph{\explanation}
\textit{Underestimation} problems, as well as the the \textit{Insufficient workforce}, are related to a larger problem of \textit{Project's Scope}.
The majority of planning problems are due to optimistic estimation: typically due to (1) tasks that developers thought easy and fast to complete and (2) the time needed to create game assets (e.g., 3D models and music).

\begin{quoting}[endtext='' -- P\#417]
Back in the days when we crafted our first budget and milestone plan we had the development of [the game] ironed out to five full-time developers working for six months. Fact: [the game] took eight full-time and between two and four part-time developers 24 months to barely finish. Our initial estimate was off by more than 700 percent.
\end{quoting}

\paragraph{\solutions} Developers believe that they must allocate time to do ``everything correctly'' to achieve a more ``solid game''. Thus, goals and deadlines must be defined early and, if needed, re-defined often during the production. Also, they must spend more time assessing risks during pre-production, allocating more time for every detail of the game, not letting anything be added as ``afterthought''.

\subsection{Unclear Game Design Vision}
\label{sec-vision}

\paragraph{\explanation} Teams often face difficulties in specifying the core mechanics of the game. They normally write a Game Design Document (GDD), defining the project and its scope, during the pre-production phase. This document is also used to divide tasks and define the artistic designs. However, writing such document is difficult and requires game development expertise. This document is also rarely updated during the projects' life and the game design visions change regardless of the definitions contained within it. Unclear game design vision is also caused by the absence of a clear playtesting process and by problems in the team, e.g., poor division of tasks and lack of brainstorming. The divergence of creative views between game designers and a publisher is also a common problem.

\begin{quoting}[endtext='' -- P\#381]
Although all of the changes we made along the way made for a better final product, the ever-changing design definitely added to development time and made it more difficult to balance the gameplay experience.
\end{quoting}

\paragraph{\solutions} A solution to achieve a clearer game design vision is to spend more time on pre-production and prototyping. It also includes investing more time playtesting the game. Also, game developers should follow traditional software-engineering processes, in particular enforcing ``feature lockdown dates'' to stop new additions to the game, do fewer review cycles to avoid staggering the workflow, and define the set of tools before going into production.

\subsection{Lack of Fun}
\label{sec-fun}

\paragraph{\explanation} Game development generally includes iterations to find and refine the game ``fun factor''. In large game projects, the core concepts of the game, including its ``fun'', are established during the pre-production phase; in indie games, these are defined during development. Yet, during production, a game may prove less fun than expected in pre-production. A game that is not fun is software without purpose. Developers must then add new features, or change existing ones, to increase the ``fun'' factor, which leads to wasted work and delays. The causes for a lack of fun vary with each game and their premises. The most common causes are weak mechanics, ugly art, unrealistic or unappealing story, and lack of tutorial for new players.

\begin{quoting}[endtext='' -- P\#123]
[The game] really is a simple game, and in some respects, it's too simple. There's no character progression, no levels, and no real incentive for the player to keep coming back.
\end{quoting}

\paragraph{\solutions} Developers recommend three steps to prevent/overcome this problem. First, they recommend more playtesting sessions to identify the weak points of the game and to survey  (early adopter) players about the game. Second, they suggest spending more time balancing the game and polishing the players' experience, for example, by reducing the players' frustrations. Third, they advise investing in tutorials to help players. Developers also advocate investing more time in prototyping during pre-production.

\subsection{Platform and Technology Constraints} \label{sec-platform}

\paragraph{\explanation} Game developers often face problems with platform and technology constraints. They must contend with different consoles, mobiles and variety of hardware/software in PCs. Consequently they must write code dedicated to manage, for example, the memory allocation, graphical details, and load times. They also must account for old technology slowing down development and difficulties when creating a multiplayer experience.

Developers routinely face problems with memory, especially when working with low-level code, ``closer'' to the hardware. They often do not consider platform constraints when designing their games, adding content regardless of these constraints, yielding long build times, long start times, and games that may not run at all on certain platforms.

\begin{quoting}[endtext='' -- P\#358]
Developing a PC game is very different from a console game, particularly in terms of memory management, loads and saves.
\end{quoting}

\begin{quoting}[endtext='' -- P\#441]
We knew [the open sandbox world] would be an issue, but we underestimated the painful impact this would have on total memory usage and we planned poorly from the start.
\end{quoting}

\begin{quoting}[endtext='' -- P\#106]
It took about five minutes to load a single level on a developers station. Therefore, it took about five minutes to test the smallest change.
\end{quoting}

\paragraph{\solutions} Developers use different solutions to overcome technical problems. However, these solutions do not generalise to different games. Common, general solutions include understanding the architecture and hardware constraints, working closely with the platform developers, e.g., console and GPU manufacturers, prioritising core game mechanics and leaving non-essential features for future updates.

\subsection{Game Design Complexity}
\label{sec-complexity}

\paragraph{\explanation} Developers often struggle with the complexity of game design. It stems from the scope of the features: ambitious features are abandoned, before the project even starts, due to lack of resources. Even when the scope is reasonable, a large numbers of features makes it difficult to follow the initial vision. Tight deadlines and parallel projects also damage the game design.

\begin{quoting}[endtext='' -- P\#58]
All of us had extremely high expectations for the game, but the total feature set turned out to be unrealistic given our small development staff and fixed schedule.
\end{quoting}

\paragraph{\solutions} Developers advise simplifying the game design: visual style, scenario, and even the game achievements\footnote{In video gaming, an achievement is a meta-goal defined outside a game's parameters.}. They also suggest carefully planning the game levels, instead of rushing into production. They recommend using better tools for the tasks and paying attention to camera misuse.

\subsection{Inadequate or Missing Tools}
\label{sec-tools}

\paragraph{\explanation} Tools rarely offer all of the features needed by developers to build their games. Developers report three main problems with tools: (1) inadequate or buggy game engines, (2) tools not fulfilling the special requirements of games, and (3) expensive tools that cannot be purchased for lack of financial resources.

\begin{quoting}[endtext='' -- P\#898]
Your game engine shapes your entire development and limits what you can and can't do. In this matter, I chose poorly.
\end{quoting}

\paragraph{\solutions} Developers advise evaluating the tools to be used during pre-production, because any change during production is financially costly. They suggest carefully assessing building their own tools or purchasing third-party software. This decision depends on the type of game.

\subsection{Misaligned Teams}
\label{sec-team}

\paragraph{\explanation} In large game companies, where many different teams work on the same game, people may establish different/diverging visions on the game design and development. In small game companies, misalignment may happen when developers cannot reach agreement regarding game design or development choices, often due to lack of dialogue or conflicting personalities.

\begin{quoting}[endtext='' -- P\#349]
(...) my style of development was in conflict with what they wanted to do. I tended to be more conservative with how I code things. I am less interested in using the latest/greatest STL techniques than I am in having readable code (...) This created some clashes during development as I would get frustrated when I'd find a bug in something that I found difficult to read. 
\end{quoting}

\paragraph{\solutions} Developers suggest that managers keep teams aligned. Meetings are important to fulfill this objective. Developers also suggest improving teams' organisations and communication, especially between art and technical teams.

\section{Related Work}
\label{sec:related}

We now summarise academic work related to problems in game development. We discuss each work individually and compare them to our own work in Section \ref{sec:related-summary}.

\subsection{\citeauthor{Callele2005}, \citeyear{Callele2005}}

\citet{Callele2005} analysed 50 postmortems from the Game Developer Magazine, written between 1999 and June 2004, and investigated how requirements engineering was applied to game development. They grouped ``What went right'' and ``What went wrong'' into five categories: (1) \emph{Pre-production}, problems outside of the traditional software development process; (2) \emph{Internal}, problems related to project management and personnel; (3) \emph{External}, problems outside of the development team's control; (4) \emph{Technology}, problems with the creation or adoption of new technologies; and, (5) \emph{Scheduling}, problems related to time estimates and overruns.

They reported that internal problems are 300\% more prevalent than that in others categories. Most internal problems relate to project management: missing tasks and poor task estimation.

\begin{quoting}[endtext='' -- \citet{Callele2005}]
Project management issues are the greatest contributors to success or failure in video game development. In the case of failure, many of these issues can be traced back to inadequate requirements engineering during the transition from pre-production to production.
\end{quoting}

We also found, in our dataset, that management problems form a large percentage of the problems, even more so if we also consider business problems as management problems. However, we also observed that another large percentage of problems happen during production and include game design problems, technical problems, and problems with tools.


The authors also reported that management problems are due to the transition between pre-production and production. In game development, developers usually use pre-production to validate game concepts through prototypes and plan features and schedule. Similarly, we found in our dataset that planning, feature creep, and, to a less degree, delays, are recurring problems during game development. However, we observed that feature creep decreased over the years.

\subsection{\citeauthor{Petrillo2009}, \citeyear{Petrillo2009}}

\citet{Petrillo2009} analysed 20 postmortems published on the Gamasutra Website to identify recurring problems and compare them with traditional software-engineering problems. They concluded that (1) video-game development suffers more from management problems rather than technical ones; (2) problems in video-game development are also found in traditional software development; and, (3) common problems are \emph{Scope}, \emph{Feature Creep}, and \emph{Cutting Features}.

They also reported that multidisciplinary teams in large game studios are also a source of problems:

\begin{quoting}[endtext='' -- \citet{Petrillo2009}]
The team in traditional software engineering is usually relatively homogeneous. However, the electronic games industry (...) attracts people with a variety of profiles such as plastic artists, musicians, scriptwriters, and software engineers.
\end{quoting}

In our dataset, we also identified many problems related to teams, even in indie studios with few developers. In particular, we reported that problems related to teams and communication remain constant over the years. Thus, we confirm the previous observations reported by Petrillo et al.

These authors also reported that requirements engineering for games differs from traditional software as it is hard to define the fun factor of the game.

\begin{quoting}[endtext='' -- \citet{Petrillo2009}]
Another important difference is that elaborating game requirements is much more complex, since efficient methods to determine subjective elements such as ``fun'' do not exist.
\end{quoting}

Similarly, we observed in our dataset that, during production, developers add new features to their games, to create better games, but against their prior requirement analyses. Although developers set the game mechanics (features) during pre-production, they often change/add new features during production, in particular to increase their ``fun''.

\subsection{\citeauthor{Kanode2009}, \citeyear{Kanode2009}}

\citet{Kanode2009} used postmortems to discuss the challenges of adapting traditional software engineering to video-game development. They reported differences between game development and traditional development, which we summarise in \autoref{tab:kanode}.

\begin{table*}[width=1\textwidth,pos=htb!]
	\footnotesize
	\caption{Challenges and Practices in game development adapted from \citet{Kanode2009}.}
	\label{tab:kanode}
	\begin{tabular}{@{}
			p{\dimexpr.15\linewidth}
			p{\dimexpr.4\linewidth}
			p{\dimexpr.4\linewidth}
			@{}}
		\toprule
		Challenge & Description & SE Practices \\ \midrule
		Diverse Assets & Increasing complexity, diversity and size of art assets. & Optimize tools and pipeline for integrating assets into the game. \\ \addlinespace
		Project Scope & Poorly established project scope further compounded by feature creep. & Keep project scope realistic and consider time for game exploration and feature creep. \\ \addlinespace
		Game Publishing & Bringing a video game to market involves a game development company convincing a game publisher to back them financially. & Better communication with the publisher, keeping requirements clear and informed project progress. \\ \addlinespace
		Project Management & The management of a game development project involves the oversight of multidisciplinary teams. & Invest in managerial training with an emphasis on project management practices. \\ \addlinespace
		Team Organization & Teams are segregated by specialty (programming, design, etc) or with functional units (combination of expertises). & Encourage an attitude of the ``team as a whole'' and less importance on individuals. \\ \addlinespace
		Development Process & The over-arching phases of game development are pre-production, production, and testing. & Understand current process and the problems with it. Identify processes that will benefit the project. \\ \addlinespace
		Third-Party Technology & Due to costs, complexity, and higher consumer expectations, game developers are using more components from third parties. & Apply risk management to selection of third-party technology in order to identify which components would work best. \\ \bottomrule
	\end{tabular}
\end{table*}

\paragraph{Asset Diversity:} They identified many technical problems with assets. These problems principally seem to be due to the large numbers of assets, not their diversities. Developers reported problems managing assets and performance, e.g., long load times.

\paragraph{Project Scope:} Problems with the games' scopes are recurring. Developers define excessive scope and time/budgets constraints force them to cut features. Feature creep is still a problem, but has decreased over the years.

\paragraph{Game Publishing:} Problems with game publishing also appear in our dataset. Developers have difficulties with one another, and with publishers, especially indie studios with little experience. However, publishers are important to the success of games, even if politics and creative interventions are often perceived to hurt game development. The relationship between developers and publishers deserves more research but is out of the scope of this work.

\paragraph{Project Management \& Team Organization:} We also observed many problems related to project management and teams, with the exception of indie game studios, in which each developer performs more than one function.

\paragraph{Development Process:} The development phase is usually split into pre-production and production. The testing phase varies depending on the game genre. For example, a multiplayer, service-based game like ``Dota 2'' will have a different testing process than a single player game like ``Hollow Knight''. In our dataset, we observed many postmortems stating that pre-production was skipped, with dire consequences during production.

\paragraph{Third-Party Technology:} Third-party tools help new developers write games. Game engines, for example, were a major contributor to the surge of indie game studios. However, problem with tools exist over the years, regardless of how advanced they are.

\subsection{\citeauthor{Lewis2011}, \citeyear{Lewis2011}}

\citet{Lewis2011} used two previous papers \cite{Blow2004, Tschang2005} to identify problems in game development and assess whether/why these could be of interest to software-engineering researchers. They highlighted some differences between games and traditional software.

They reported that, in large game studios, \textit{teams} are multidisciplinary and tightly coupled and that they suffer from tight budgets and deadlines. They also wrote that larger teams require strong leadership due to constant developer turnover. We found similar problems in our dataset. We observed different problems for smaller or indie game studios: small studios do not have the budget to build new game engines, which constrains their workflows.

Regarding \textit{tools and environments}, the authors reported a lack of quality tools. We also identified problems with tools. They based their report on the lack of tools to handle the complexity of game development. In our dataset, we also found discussions about game engines, which, although not without flaws, may ease game development.

The authors discussed the lack of \textit{design patterns} for game development. The information from postmortems does not give us this level of granularity but showed problems with game design and technical aspects of the games.

The authors referred to game engines as \textit{middleware} that facilitates game development. They also stated that engines often need rewriting to provide the features needed by the developers. In our dataset, we also noticed that developers struggle to implement features because of the game engine or some other technological choices.

The authors classified games as emergent software, for which we cannot predict the outcome. They mentioned that game studios prefer to hire dozens of human testers instead of using unit tests. We concur with these statements: in our dataset, only one problem is related to unit testing. All other testing problems refer to playtesting sessions with players.

\begin{quoting}[endtext='' -- \citet{Lewis2011}]
(...) digital game designers have tried to design a game upfront through copious amounts of documentation, but that the documentation is made instantly obsolete by surprises that arise when actually implementing the game.
\end{quoting}

Finally, the authors stated that documenting a game upfront is pointless as new features are added regularly, making documentation obsolete. We observed only 2\% of documentation problems in our dataset. Some developers stated the need for a clear vision, but not game-design documents. Developers want a clear vision more than documentation.

\subsection{\citeauthor{Washburn2016}, \citeyear{Washburn2016}}

\citet{Washburn2016} analysed 155 postmortems, written over 16 years, and identified some characteristics and pitfalls of game development, and suggested good practices. They divided problems and practices into five categories (\textit{Product}, \textit{Development}, \textit{Resources}, \textit{Customer Facing}, and \textit{Other}) and 21 sub-categories. They discussed four of the most common problems, shown in \autoref{tab:washburn}, adapted from \citep{Washburn2016}.

\begin{table*}[width=1\textwidth,pos=htb!]
	\footnotesize
	\caption{Main findings of \citet{Washburn2016}.}
	\label{tab:washburn}
	\begin{tabular}{@{}
			p{\dimexpr.1\linewidth}
			p{\dimexpr.2\linewidth} 
			p{\dimexpr.04\linewidth}
			p{\dimexpr.22\linewidth}
			p{\dimexpr.34\linewidth}@{}}
		\toprule
		Category & Description & \% & Details & Takeaway \\ \midrule
		Game Design & Good or bad design decisions that impacted the quality of their game & 22\% & Overly ambitious game designs which could not be implemented and concepts that confused the player & Keep implementation in mind while creating a design, and create contingencies if it cannot be done. Test key game concepts before release (audience reception) \\
		Dev. Process & The process teams use while developing affects the quality of the product & 24\% & Developer did not plan before the development and also mismanagement & To avoid conflicts during the development process, teams need to have proper management and invest time upfront planning before beginning development \\
		Obstacles & Obstacles are more likely to have a negative impact on a team & 37\% & Lack of team dynamic and unfamiliarity among the team & Developers should participate in team building. Subscribe to a method of risk management, because they are more likely to face obstacles than more seasoned teams \\
		Schedule & Missed milestones or delivered them late & 25\% & Problems in estimation, optimistic scheduling, and design changes late in development & To avoid schedule slippage, developers need to spend more time to plan out all the work that needs to be done so that no tasks are overlooked when giving estimates \\ \bottomrule
	\end{tabular}
\end{table*}

The authors reported that the most common problem relates to \textit{Teams}, similar to our observations in which \textit{Teams} problems are the third most common (8\%), e.g., lack of communication and disagreement among developers.

They also reported that scheduling and process are recurring problems, which we also support with our findings from the dataset: underestimation and management are reported as the main causes of planning problems.

The authors cited ambitious scope and confusing concepts as examples of game design problems, which we also support via our findings, although we found a more diverse set of game design problems.

\subsection{\citeauthor{Edholm2017}, \citeyear{Edholm2017}}

The authors conducted interviews at four different game studios and reviewed 78 postmortems to investigate the culture of crunch-time in the game industry. According to their interviewees, crunch time is common within the game industry as the majority of game studios applied such practice.

\begin{figure}[htb!]
\includegraphics[width=1\linewidth]{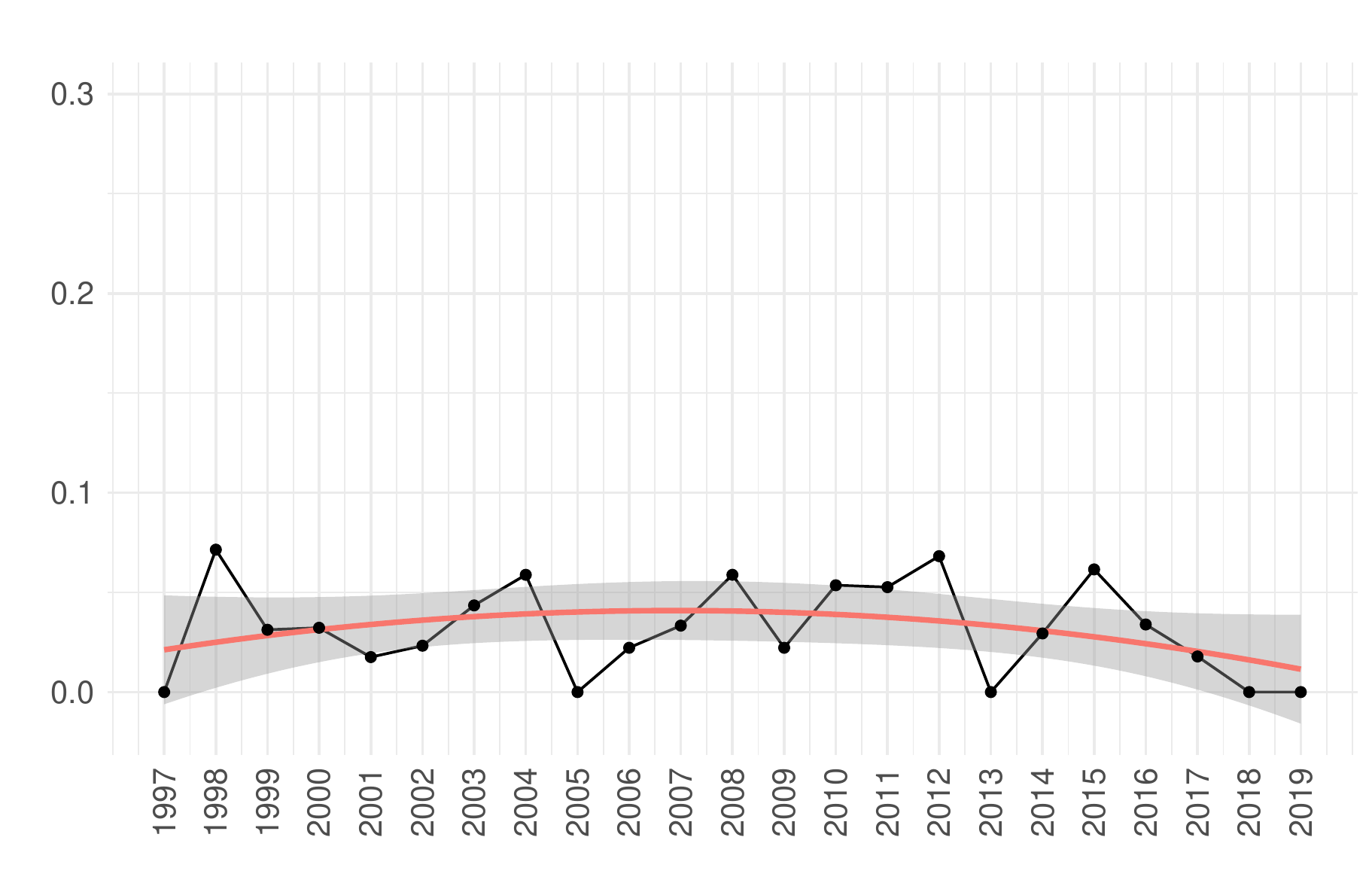}
\caption{Crunch-time problems}
\label{fig:historical-patterns-crunch}
\end{figure}

From their postmortem data, 45\% mentioned crunch-time. Also, crunch-time has been within game industry from early 2000 to the current date (2014). Moreover, small studios are more prone to crunch (54\% crunch) than both micro-(33\%) and medium-sized (36\%) studios.
Our data showed the first signs of crunch-time in 1998, but it is decreasing after 2015, with zero mentions in 2018 and 2019 (\autoref{fig:historical-patterns-crunch}).

\begin{quoting}[endtext='' -- \citet{Edholm2017}]
(...) well-being of the product is prioritised over employee welfare. Since people have a personal investment in the product they create, they blame themselves if it ends up badly.
\end{quoting}

The authors also investigated other problems and found that Planning and Technical problems are common while publisher disagreement, pressure, and un-fun game are rarely mentioned. Our data show similar results. Planning is the fifth more common problem and technical is the first one. However, we found many problems related to publishers like \textit{communication}, \textit{planning}, and \textit{marketing}. \textit{Lack of fun} is also one of the most common root causes in our dataset.

\subsection{Summary}
\label{sec:related-summary}

\autoref{tab:related-works} summarises these previous works, their methods and goals. These previous works used postmortems to discuss video-game development problems. They used \textit{ad-hoc} classifications for the problems. When comparing their findings with our data, they concur (with some exceptions). 

Management is found to be the main problem in game development by all these works and ours, possibly because the source of information are postmortems, in which senior developers are more willing to discuss ``general'' problems rather than technical minutiae. Previous works did not consider indie studios or game engines. 

\begin{table}[pos=htb!]
\footnotesize
\centering
\caption{Summary of the related works.}
\label{tab:related-works}
\begin{tabular}{@{}llrl@{}}
\toprule
\multirow{2}{*}{Paper} & \multicolumn{2}{l}{Postmortems’}      & \multirow{2}{*}{Study goal} \\ \cmidrule(lr){2-3}
                       & Analysis & \multicolumn{1}{r}{\#} &                             \\ \midrule
\citeyear{Callele2005} \citet{Callele2005}                & Yes      & 50                         & Requirements                \\
\citeyear{Petrillo2009} \citet{Petrillo2009}               & Yes      & 20                         & Problems                    \\
\citeyear{Kanode2009} \citet{Kanode2009}                 & Yes      & ?                          & Challenges                  \\
\citeyear{Lewis2011} \citet{Lewis2011}                  & No       & --                         & Problems                    \\
\citeyear{Washburn2016} \citet{Washburn2016}               & Yes      & 155                        & Characteristics             \\
\citeyear{Edholm2017} \citet{Edholm2017}                 & Yes      & 78                         & Crunch-time                 \\ \midrule
This paper             & Yes      & 200                        & Problems                    \\ \bottomrule
\end{tabular}
\end{table}


\section{Follow-up Discussions}
\label{sec:discussion}

We now discuss the ten problems showed in \autoref{sec:res-subtypes} and present suggestions to address each one.

\subsection{Insufficient Workforce}

We observed that \textit{Insufficient workforce} is mainly caused by poor management of the project \textit{Scope} (requirements), which leads to other problems like \textit{Cutting Features} and \textit{Crunch Time}. Mitigation depends on the game company. Game developers report that pair programming \cite{Kude2019} and code reviews \cite{Alami2019} are not common in game industry while they are well-established practices in traditional software development. Similar practices could be adapted in the game industry, even pairing technical and non-technical developers.

Less common than \textit{insufficient workforce} is \textit{overstaffing}, which happens when a game company assigns too many developers to a given task, making communication and organisation difficult. \textit{Overstaffing} is also the result of poor estimation, often when managers and stakeholders interfere with one another. Despite developers' complaints about staff shortage, game companies area advised to remember Brooks' Law \cite{Brooks1978}.

\begin{quoting}[endtext='' -- P\#518]
When there's so much work to do, one of the first reactions is to throw more hands into the mix thinking that this will lighten the load. (...) Rather, we found that, an increasing the number of people on the team to aid the workload inhibited the project, and placed great strain on the lines of communication.
\end{quoting}


\begin{suggestionbox}
Problem: Insufficient workforce (Section~\ref{sec-understaffing})
\tcblower
Consider ``sharing the load'' of complex tasks, avoid blindly hiring more workforce (e.g., Brooks' Law), and consider outsourcing if appropriate.
\end{suggestionbox}

\subsection{Environmental Problems}




Environmental problems relate to managing people, like in any other tech.\ company. Common sense dictates following best practices, e.g., balancing developers' expertise levels and having small teams with small scopes. Game development also has some specific characteristics, e.g., crunch times and mistreatment. These problems require protecting developers and allowing them to perform their job properly.

\begin{suggestionbox}
Problem: Environmental problems (Section~\ref{sec-environment})
\tcblower
Balance expertise levels among developers, keep small teams with small scopes, and shield developers from external interference.
\end{suggestionbox}

\subsection{Wrong Marketing Strategy}

Marketing is the problem type that increased the most in the study period. We observed that its main causes are threefold: new audience acquisition (need for new strategies), lack of expertise in promoting games (especially in indie companies), and saturation of the game market (need to stand out).

The way developers communicate about their games e\-vol\-ved from magazines in 1997, through forums, social media, online stores, to today's streamers and independent reviewers\footnote{\url{https://bit.ly/2Zyu77Q}}. Twitch\footnote{\url{https://www.twitch.tv/}}, the most popular streaming platform, has 1.645 billion hours watched per month\footnote{\url{https://bit.ly/3h2uTQ8}}. The most popular, independent, YouTuber has millions of views per day\footnote{\url{https://bit.ly/393kZuZ}}. Large outlets, e.g., Polygon and IGN, still play an important role but streamers can increase sales dramatically\footnote{\url{https://bit.ly/3j7nBg9} and \url{https://bit.ly/3jdyltp}}.

Taking advantage of new media is difficult. Indie developers, with low marketing budgets, often fail by trying to reach too many ``influencers''. A lack of marketing expertise and a crowded game market\footnote{\url{https://bit.ly/30fTUAI}} make it difficult to be noticed. 

The game market demands that developers have a more transparent relationship with players, like streamers. Developers should create awareness, promoting the strengths of their games, building relationships with players (using alpha/beta testing, answering constructive feedback, creating development blogs), and outsourcing marketing, if needed.

\begin{suggestionbox}
Problem: Wrong marketing strategy (Section~\ref{sec-marketing})
\tcblower
Do not promote games naively; focus on their strong points, and build/strengthen relationships with users.
\end{suggestionbox}

\subsection{Underestimation}

Hofstadter's Law states that ``It always takes longer than you expect, even when you take into account Hofstadter's Law" \cite{Hofstadter1999}, which holds true for game development. Software estimation is a well studied field, which uses previous data to estimate the effort (and cost) of a project \cite{SWEBOK2014}. It uses different methods, e.g., COSMIC or Agile' Story Points, and techniques, e.g., experts' analysis, machine learning \cite{Nassif2016}, etc. 

Yet, software estimation in game development is often performed manually, using  senior developers' experience. Developers, especially senior ones, aside from using their experience, should also document their experience for future use, e.g., for creating ML models.

Game estimation, like for any other software projects, varies across game projects. Teams move from one game to another and must adapt to technological advances and different requirements, which make estimation difficult. They must invest in long pre-production phases to research and understand new technologies, tools, and game designs.

We believe that better estimation comes with better information about the previous projects. However, the closed nature of most games makes sharing information difficult. Postmortems are an important source of information, yet insufficient \cite{Politowski2018}. Developers should gather metadata about past game projects and apply/extend traditional estimation methods. For example, although the COSMIC method is ``technology independent'' \cite{Sadowski2019}, it should be adapted to game projects and their unique aspects (e.g., art assets).

\begin{suggestionbox}
Problem: Underestimation (Section~\ref{sec-understimation})
\tcblower
Avoid relying solely on human expertise and invest in building a knowledge base about past projects to better estimate future games.
\end{suggestionbox}

\subsection{Unclear Game Design Vision}

An unclear game-design vision impacts the entire game project, including management and testing. Although related to game design and art, the game-design vision must be embraced by the whole team and, thus, is also a management problem. 

Teams need to understand the project vision to avoid wasted work. They should spend less time defining static documents, which become quickly obsolete, and more time in pre-production until the core mechanics and the fun factor is clear. They must prototype and playtest. Finally, they should keep creative control over the projects, to the greatest extent feasible.

\begin{suggestionbox}
Problem: Unclear game-design vision (Section~\ref{sec-vision})
\tcblower
Keep a clear vision of the game design. Do not waste time with static documents. Spend more time prototyping and playtesting.
\end{suggestionbox}

\subsection{Lack of Fun}

Lack of fun is essentially a game-design problem, out of the scope of software engineering. Research about ``fun'' in games is related to the study of human cognition and also to the human capacity/curiosity to learn new things \cite{Koster2013}.

Developers should allocate time to polish the mechanics, art, and story of their games. They should use playtesting (and players' surveys) to identify weak points in their games.

\begin{suggestionbox}
Problem: Lack of fun (Section~\ref{sec-fun})
\tcblower
Allocate time to find the ``fun'' in the game mechanics. Polish art and story. Extensively playtest games to identify their weak points.
\end{suggestionbox}

\subsection{Platform and Technology Constraints}

The gaming market is spread across different platforms and developers must publish on different platforms to reach more players and sell more games. Platform constraints stem from the differences among/within consoles, mobiles, and PCs. Platform constraints are often defined by the lowest common denominator in consoles, mobiles, and PCs. They include slow read-and-seek times on hard drives, CPUs with low clock speeds and numbers of cores, and old graphics cards.

Developers must understand each platform and developing for multiple platforms adds complexity to game projects. Yet, even with multi-platform game engines, they must deal with each platform's constraints and must degrade their games to accommodate the lowest common denominator (particularly frame rate).

Moreover, consoles are built to sell in quantity and must be cheap and, thus, often use ``outdated'' technologies\footnote{This trend seems to come to an end with the new XBox series X and PS5 consoles with SSDs and GPUs.}. Yet, developing for consoles has benefits: developers do not need to handle a large variety of hardware configurations and can offer the same game play to all players.

Developers should assess the viability of their games on the technical specifications of the target platforms (for which they should reserve time for experimentation).They should also \textit{gracefully degrade} their games on lower-end devices or \textit{progressively enhance} them on more capable platforms. 

\begin{suggestionbox}
Problem: Platform and technology constraints (Section~\ref{sec-platform})
\tcblower
Have a better understanding of the target platforms (architectures and limitations) before committing to a game project.
\end{suggestionbox}

\subsection{Game Design Complexity}

Game design is difficult and there is no clear process to follow. We observed that a ``good'' game design requires: (1) a clear game-design vision, something on which all developers agree; (2) a deep understanding of the players' expectations; and, (3) constant playtesting sessions to evaluate and validate the games.

Keeping the game design simple, focusing on players, and continuous playtesting can mitigate issues with game-design complexity. Building a relationship with the players can also improve the game through quality feedback.

\begin{suggestionbox}
Problem: Game design complexity (Section~\ref{sec-complexity})
\tcblower
Keep the game design simple, pay attention to players' expectations, adopt continuous playtesting, and obtain feedback from the players.
\end{suggestionbox}

\subsection{Inadequate or Missing Tools}

Tools often frustrate developers, in particular game engines. For example, in two game projects, EA\footnote{Electronic Arts is a publisher and owner of many video game studios, see \url{https://www.ea.com/en-ca}.} forced their developers to use their proprietary Frostbite engine, causing delays and reworks for the game ``Dragon Age 3'' \cite{Schreier2017} and the failed project ``Anthem''\footnote{\url{https://bit.ly/39fTnTt}}.

Game engines can speed up game development but also constrain game designs. They are few, including Unity and Unreal, and proprietary, closed-source engines in large companies. Although open source, Godot\footnote{\url{https://godotengine.org/}} is not yet as mature as its proprietary counterparts \cite{Politowski2020}. Developers may consider building their own game engine but should carefully assess the benefits and the risks of doing so.

Game developers should carefully choose their game engines according to: (1) the project goal -- Can we implement the game using this engine? (2) the team experience -- Is the team comfortable with this engine? (3) the development schedule -- Should we build, extend, or use a third-party engine? and, (4) the game budget -- What is the trade-off between licensing and supporting our own game-engine?

\begin{suggestionbox}
Problem: Inadequate or missing tools (Section~\ref{sec-tools})
\tcblower
Allocate time to experiment with multiple tools and game engines before choosing/building one.
\end{suggestionbox}

\subsection{Misaligned Teams}

Misaligned teams arise from communication problems. In large companies, teams are usually divided into design, art, and technical departments. Keeping them aligned with one another and with the game-design vision is a management problem. Designers define the goals of the games. Artists realise the game design. Managers keep the teams cohesive. They all benefit from having the same vision. Also, they all should be aware of the details of the games: next steps, colleagues' roles, etc. 

Game studios should consider applying agile practices, reducing teams sizes, and easing communication among teams. Instead of dividing teams by departments, managers should consider mixed, independent teams, that can design, implement, test, and integrate features into games. The design process should also change to allow for modular games.

\begin{suggestionbox}
Problem: Misaligned teams (Section~\ref{sec-team})
\tcblower
Consider changing the team structure/process if there are communication problems among departments/teams.
\end{suggestionbox}


\section{General Discussions}

We now take a step back and discuss other important problems with game development.

\subsection{Problems Evolution}

\textit{Production} problems remain constant to today. The most clear spike in the data occurred in 2005, which might be related to the arrival of a new console generation that year: the seventh generation, e.g., Sony Playstation 3, was released between 2005 and 2006. The Playstation 3, with its new architecture\footnote{\url{https://venturebeat.com/2014/07/06/last-gen-development/}}, was notoriously difficult to program and Sony shared some information only with first-party studios\footnote{\url{https://cnet.co/3haIEfN}}.

\textit{Management} problems peaked in 1998 and are less frequent now. One factor that helped decrease management problems might be the adoption of agile methods. The game industry, even today, often works with old development methods, e.g., Waterfall \cite{Politowski2016}, yet agile methods, born around the 2000s, are being adopted gradually. Pre-production and Production may not be both amenable to agile methods. However, the concrete development of the game, during production, could benefit from using agile methods. We discuss further development methods in Section \ref{Section: Development Processes}.

The problems with \textit{Business} increased over the years. Our hypothesis is that the rise of Indie developers, in particular the ``one-man-army'' teams in which one developer does all the tasks\footnote{Some examples of (successful) games written by only one developer are ``Stardew Valley'' and ``Dust: An Elysian Tail''.}, contributed to increasing this problem. Indie developers do not have publishers or colleagues to deal with marketing and often perform related tasks poorly. Their business knowledge is often limited.

\subsection{Development Processes}
\label{Section: Development Processes}

The game industry still mostly uses waterfall-like development processes \cite{Politowski2016}. Their processes usually divide into: 
(1) \textit{Pre-production} to prototype, find the game's core ``fun'' mechanic, create the Game Design Document (GDD), etc. This step is normally done with a small team of developers. 
(2) \textit{Production} to implement the game with the full team. The implementation process varies according to the team. 
(3) \textit{Post-production} to work on updates and bug fixes.
The traditional iterative process may be adequate for game projects.
However, constant evaluation of team productivity will aid the team to track progress and help in the decision to hire more developers.



\subsection{Developer Turnover}

``The game industry is cyclical, constantly churning employees in and out depending on the needs of a project''\footnote{\url{https://bit.ly/2WmLWET}} and, thus, game developers often change companies. Therefore, teams are also constantly changing, having to adapt to newcomers. This turnover happens during all the development phases, because game projects are long duration projects. Even with a clearly-defined process for newcomers, with mentoring from senior developers, their productivity will be low at first, yet micromanagement must be avoided\footnote{The director of Final Fantasy 14 had to micromanage the team to keep the production pace, but he advises not to do it in \url{https://youtu.be/Xs0yQKI7Yw4}.}.

\subsection{Game Testing}

Given the importance and emphasis given to software testing in traditional software development, we were surprised to find little information about testing in game projects in the postmortems. 
One hypothesis could be that testing is largely successful and therefore does not need mentioning in the postmortems. However, it is well known that games often suffer from low quality and that game projects often overrun their schedules, hinting that testing is probably problematic.
Therefore, our next hypothesis is that testing, in particular software-engineering testing, is under-performed by game developers. Indeed, postmortems mention playtesting but do not mention unit testing or integration testing. This lack of mention is interesting and calls for more research on game developers' testing habits (or lack thereof) and the reasons for these habits.

Video-game testing, in practice, includes the gameplay testing sessions. Game testers have the duty to spot problems, reproduce previous problems, and assess the gameplay mechanics (among other measurements that the game designers must verify). The techniques used to test games are mainly manual, relying on the \textit{ad-hoc} reasoning of game testers as they play the game again and again, obtaining extensive knowledge about the games.

Playtesting sessions are not scalable and none of the studied postmortems mentioned automation. Therefore, future works could study (semi)automated playtesting. For example, automation of regression testing can reduce the load from human testers. Also, deep learning could be used to create agents to explore games \cite{Zheng2019}.


\section{Meta Discussions}

We now consider our study in itself as a object of study through threats to its validity and meta-discussions.

\subsection{Threats to Validity}
\label{sec:threats}

We presented an analysis of the gray literature on game-development problems found in game postmortems.

\paragraph{Dataset based only on postmortems:} Our results are based \emph{only} on postmortems, which do not represent all the games or the whole game industry. Nonetheless, postmortems are the best (and only) source of information to which we have access, i.e., publicly available.

\paragraph{Dataset only has successful projects:} All the postmortems were gathered from the Gamasutra Web site. They pertained to 200 game projects that were released and were, for most of them, profitable. Therefore, they did not include failed game projects, which may lead to optimistic conclusions \cite{Petrillo2009}. Yet, they all reported problems, which we identified and analysed in this work. 

\paragraph{Developers might not tell the whole story:} As shown by \citet{Washburn2016}, some authors of postmortems may not disclose all that happened during their game projects. Thus, postmortems do not represent the entire reality. Yet, they provide list of meaningful problems to which (1) game studios should pay attention in their own, next projects and (2) researchers should investigate to find solutions.

\paragraph{Biases introduced by the developer role:} The awareness of the situation covered in the postmortems might be biased by their author's role within the organization. Sometimes, the postmortems were written by two or more developers, with different roles (programmers and designers, for example). For others, only one person wrote each postmortem. Moreover, there were many different authors' roles among all the 200 postmortems in the dataset. Regardless, all postmortems were written by at least one developer who belonged to the development team. Also, postmortems are the only source of information about game development. Therefore, although possibly biased, we considered them useful.

\paragraph{Problems are too abstract:} For lack of space and difficulty to convey the whole context, the problems described by the authors of the postmortems are abstract, often without technical details. We identified and formalised the problems from free texts, sometimes written by designers or managers unaware of the technicalities faced by game developers. Yet, the diversity of authors of the postmortems is valuable and reduces any bias towards one particular game studio or one particular game genre. Also, they provide a more complete view of the problems that they faced, including problems with management, design, marketing, etc.

\paragraph{Research bias:}
The analysis of problems relied on our own interpretation of the postmortems and the reported problems. This interpretation could vary according to each researcher. To reduce any bias, we discussed the problems and our interpretations in each iteration of reading the postmortems, updating the catalogue of problem types only when necessary, until we reached a fixed point, as shown in \autoref{fig:method}.

\paragraph{Different numbers of problems per year:} At first, we chose postmortems randomly but some years have more postmortems than others so we mitigated this imbalance by dividing the numbers of problems by the numbers of postmortems per year for the historical analysis.

\paragraph{Suggestions:} Our suggestions are based not only on the data but also on our understanding of the literature, game development in particular, and knowledge about software engineering in general. Some of the problems are too specific to one project to be generalised. Therefore, we tried to be general yet avoid being obvious.

\subsection{Limitations and Solutions}

The gray literature provides invaluable insights into real problems and can complement academic works. In some cases, like this analysis of postmortems, it provides the only available source of information on a problem. However, that very benefit is a main limitation of gray literature \cite{https://doi.org/10.1111/jebm.12266}.

Indeed, gray literature may provide a single type of information (e.g., postmortems) from a single source of information (e.g., Gamasutra portal) and, thus, lack diversity. This lack of diversity is also found in the content \emph{per se} of the gray literature (e.g., only successful games and only one developer's point of view).

From this lack of diversity stem three additional, complementary problems: (1) a lack of completeness, i.e., the gray literature may provide only a partial view on a topic, (2) a lack of accuracy, i.e., the gray literature may contain documents mistakenly or purposefully misrepresenting the truth, and (3) a lack of verifiability, i.e., the content of the gray literature may be difficult/impossible to verify independently.

While these problems are unavoidable when an analysis of the gray literature stands alone, they are acceptable when the analysis serves as a starting point for more research. For example, the problems of accuracy and verifiability can be overcome through interviews and surveys with the authors of the gray literature and--or others with similar credentials as these authors. Academic research can also build on the gray literature to circumscribe a research tropic and, therefore, overcome any possible incompleteness.

Thus, for example, we propose in the next section as future work to conduct interviews with game developers. Such interviews would allow us to obtain more details on the problems faced by game developers as well as verify the given information and identify any missing information.

\section{Conclusion}
\label{sec:conclusion}





Little is known of the problems faced by game developers during their projects as the game industry has a closed-source nature. 
We used postmortems to overcome this barrier and better understand the problems of the game industry. We analyzed more than 200 postmortems, comprising \totalProblems~problems, divided in \totalTypes~types, from 1997 to 2019.

Through our analysis, we described the overall landscape of game-industry problems in the past 23 years and how these problems evolved over the years. 
We reported the following main \emph{findings}:

\begin{itemize}
    \item Based on the number of problems groups and types, the game industry suffer from \textit{management} and \textit{production} problems in the same proportion. However,  \textit{production} problems are concentrated mostly in \textit{technical} and \textit{design} while \textit{management} problems spread across all problems types.
    \item Based on the evolution of problem groups, \textit{management} problems decreased, \textit{business} problems increased, and \textit{production} problems stayed constant;
    \item Based on the evolution of the problem types:
    \begin{itemize}
        \item \textit{Technical} and \textit{game design} problems are decreasing, the latter only in the last decade;
        \item Problems related to the \textit{team} increased over the last decade;
        \item \textit{Marketing} problems had the largest increase compared to other problem types;
    \end{itemize}
    \item Finally, considering problem sub-types, the main root causes are related to people, not technologies.
\end{itemize}

In \autoref{sec:discussion} we took the liberty to give suggestions based on our findings. Therefore, this paper also presents a collection of 10 suggestions for the problems we analyzed more thoroughly in \autoref{sec:res-subtypes}.

Finally, our findings show that many problems require project-specific solutions that are hard to generalize. However, we hope that our discussion about these problems, and the suggestions, will help practitioners and researchers better understand the game industry.

In future work, we will study more postmortems to further enrich our analysis. 
We will also reach out to video-game developers to vet and further refine the identified types of problems as well as to survey their opinions on the identified solutions and our proposed suggestions. We wish to start a conversation between academia and the video-game industry, on their challenges and possible solutions.

\section*{Acknowledgement}

The authors were partly supported by the NSERC Discovery Grant and Canada Research Chairs programs.

\bibliographystyle{cas-model2-names}
\bibliography{main.bib}

\clearpage
\appendix
\section{Tables with all the problems \textit{sub-types}}
\label{sec:app-subtypes}

\begin{table}[pos=ht]
\centering
\footnotesize
	\caption{Types and Sub-types of the \textit{group} Production.}
	\label{tab:subtypes-prod}
	\begin{tabular}{@{}lll}
		\toprule
		Type & SubType & N \\ \midrule
		Bugs & Lack of proper organization/tracking & 14 \\
		Bugs & Graphic/sound issues & 9 \\
		Bugs & Platform/hardware issues & 6 \\
		Bugs & Game mechanic/system issue & 4 \\
		Design & Unclear game design vision & 28 \\
		Design & Lack of fun & 27 \\
		Design & Game design complexity & 23 \\
		Design & Balancing issues & 19 \\
		Design & Lack of polish & 13 \\
		Design & Game too short/simple & 10 \\
		Design & Release/censorship issues & 8 \\
		Documentation & Lack of design documentation & 8 \\
		Documentation & Lack of technical documentation & 7 \\
		Documentation & Poor assets management & 4 \\
		Documentation & Documentation management issues & 3 \\
		Prototyping & Not enough time or focus & 7 \\
		Prototyping & Prototype is too simple & 5 \\
		Prototyping & Prototype is too complex & 4 \\
		Prototyping & No prototyping & 3 \\
		Technical & Platform and technology constraints & 24 \\
		Technical & Optimization and Performance & 17 \\
		Technical & Game engine and Libraries & 13 \\
		Technical & Network and Multiplayer & 9 \\
		Technical & Re-work and Wasted work & 8 \\
		Technical & Build and Load time & 8 \\
		Technical & Animation and 3D & 7 \\
		Technical & Source control and file management & 6 \\
		Technical & Porting issues & 6 \\
		Technical & Networking complexity & 6 \\
		Technical & Programming language and Algorithms & 5 \\
		Technical & Production pipeline & 5 \\
		Technical & Physics and Collision & 5 \\
		Technical & Novelty and change & 5 \\
		Technical & Performance issues & 4 \\
		Technical & Patch strategies and Infrastructure & 4 \\
		Technical & Misc: UI and Localization & 4 \\
		Technical & Coding/architecture issues & 4 \\
		Testing & Insufficient test coverage & 13 \\
		Testing & Process and testing plans issues & 13 \\
		Testing & Specific project requirements & 7 \\
		Testing & Scope too big to test properly & 7 \\
		Testing & Poor feedback & 5 \\
		Testing & Reproducibility of bugs & 2 \\
		Tools & Inadequate or lack of tools & 22 \\
		Tools & Lack of expertise with the tool & 12 \\
		Tools & Concurrent tool development & 11 \\
		Tools & Middleware issues & 10 \\
		Tools & Maintenance issues & 9 \\
		Tools & Third-party issues & 7 \\
		Tools & Hardware compatibility issues & 5 \\
		Tools & Tool switch & 4 \\ \bottomrule
	\end{tabular}
\end{table}

\begin{table}[pos=ht]
    \centering
    \footnotesize
	\caption{Types and Sub-types of the \textit{group} Management.}
	\label{tab:subtypes-man}
	\begin{tabular}{@{}lll@{}}
		\toprule
		Type & SubType & N \\ \midrule
		Communication & Misaligned teams & 22 \\
		Communication & Poor dev/pub communication & 10 \\
		Communication & Poor PR & 7 \\
		Communication & Different physical locations & 3 \\
		Communication & Help/support issues & 2 \\
		Crunch-time & Not enough workforce & 7 \\
		Crunch-time & Management/financial issues & 6 \\
		Crunch-time & Growing scope & 6 \\
		Crunch-time & Publisher set tight deadlines & 5 \\
		Crunch-time & Delays/scheduling issues & 5 \\
		Delays & Technical/platform issue & 11 \\
		Delays & Poor resource management & 11 \\
		Delays & Publishing/business issues & 4 \\
		Delays & Lack of workforce & 4 \\
		Team & Insufficient workforce & 49 \\
		Team & Environment problems & 48 \\
		Team & Unexpected team disruption & 11 \\
		Team & Outsourcing issues & 5 \\
		Team & Inexperienced staff & 5 \\
		Team & Overstaffing & 2 \\
		Budget & Difficulties with external funding & 10 \\
		Budget & Limited self funding & 7 \\
		Budget & Poor management & 4 \\
		Cutting features & Not enough time & 9 \\
		Cutting features & Idea was considered overambitious & 6 \\
		Cutting features & Technical limitations & 4 \\
		Feature-Creep & Design increments over time & 10 \\
		Feature-Creep & Design increments over time & 10 \\
		Feature-Creep & Complexity of game mechanics & 8 \\
		Feature-Creep & Complexity of game mechanics & 8 \\
		Feature-Creep & Poor feature planning & 4 \\
		Feature-Creep & Poor feature planning & 4 \\
		Multiple-projects & Resource conflict & 10 \\
		Multiple-projects & Project was part-time job & 2 \\
		Multiple-projects & Procrastination & 2 \\
		Multiple-projects & Building engine at the same time & 2 \\
		Planning & Underestimation & 34 \\
		Planning & Ignoring or changing the plan & 14 \\
		Scope & Overambitious scope & 15 \\
		Scope & Poor resource estimation & 8 \\
		Scope & Lack of initial design definitions & 6 \\
		Scope & Poor scope management & 3 \\
		Scope & Poor complexity estimation & 3 \\
		Security & Piracy & 2 \\ 
\bottomrule
	\end{tabular}
\end{table}

\begin{table}[pos=ht]
    \centering
    \footnotesize
	\caption{Types and Sub-types of the \textit{group} Bussiness.}
	\label{tab:subtypes-bus}
	\begin{tabular}{@{}lll@{}}
		\toprule
		Type & SubType & N \\ \midrule
		Marketing & Wrong marketing strategy & 35 \\
		Marketing & No plan, budget, or not enough marketing & 15 \\
		Marketing & Publisher/platform/hardware problems & 11 \\
		Marketing & Game hard to market & 9 \\
		Monetization & Wrong monetization model & 9 \\
		Monetization & Game did not profit & 7 \\
		Monetization & Publisher/platform/market issues & 4 \\
		Monetization & Payment service issues & 3 \\
		Monetization & Lack of business expertise & 3 \\ \bottomrule
	\end{tabular}
\end{table}

\end{document}